\documentclass[]{pasj02}
\usepackage[switch,mathlines]{lineno} 

\usepackage{graphicx,color}
\usepackage{xcolor}
\usepackage{xspace}
\usepackage{booktabs}
\usepackage{bm}
\usepackage{ulem}
\usepackage{natbib}

\usepackage{url}
\DeclareRobustCommand{\mytt}[1]{\ifmmode\text{\path{#1}}\else\path{#1}\fi}

\usepackage[colorlinks,bookmarks]{hyperref}
\definecolor{linkblue}{rgb}{0,0,0.8}
\definecolor{linkgreen}{rgb}{0,0.5,0}
\hypersetup{linkcolor=linkblue, citecolor=linkgreen, urlcolor=linkblue}

\definecolor{valecol}{rgb}{0,0.5, 1.}
\newcommand{\vale}[1]{\textcolor{valecol}{#1}}

\newcommand{\better}[1]{\textcolor{red}{\textbf{#1}}}

\newcommand{\dif}{{\rm d}}

\newcommand{\pz}{photo-$z$\xspace}
\newcommand{\sz}{spec-$z$\xspace}
\newcommand{\zml}{z_{\rm ml}}
\newcommand{\sml}{\sigma_{\rm ml}}
\newcommand{\rml}{r_{\rm ml}}
\newcommand{\zphot}{z_{\rm phot}}
\newcommand{\zspec}{z_{\rm spec}}
\newcommand{\zbest}{z_{\rm best}}

\jyear{2026}
\Received{}
\Accepted{}

\begin{document}

\title{Optimizing the extraction of information from redshift probability distribution functions}

\author{
  Rodrigo~\textsc{Duarte},\altaffilmark{1,}\altaffilmark{2}\altemailmark \email{rodrigoduartefis@gmail.com}
  and
  Valerio~\textsc{Marra},\altaffilmark{3,}\altaffilmark{4,}\altaffilmark{5,}\altaffilmark{2}\altemailmark \email{valerio.marra@me.com}
}

\altaffiltext{1}{PPGFis, Universidade Federal do Esp\'irito Santo, 29075-910, Vit\'oria, ES, Brazil}
\altaffiltext{2}{Laborat\'orio Interinstitucional de e-Astronomia, 20921-400, Rio de Janeiro, RJ, Brazil}
\altaffiltext{3}{Departamento de F\'isica, Universidade Federal de Ouro Preto, 35400-000, Ouro Preto, MG, Brazil}
\altaffiltext{4}{INAF -- Osservatorio Astronomico di Trieste, via Tiepolo 11, 34131 Trieste, Italy}
\altaffiltext{5}{IFPU -- Institute for Fundamental Physics of the Universe, via Beirut 2, 34151, Trieste, Italy}


\maketitle

\begin{abstract}
Photometric redshifts are essential for large-scale structure analyses, yet extracting optimal point estimates and reliability measures from the probability distribution functions (PDZs) delivered by modern photo-$z$ pipelines remains an open challenge.
We introduce \mytt{turboPDZ}, a machine-learning framework that uses PDZ-derived features and associated observables to obtain an optimized redshift point estimate and a data-driven reliability score.
We apply the framework to PDZs from the three independent pipelines (DEmP, DNNz, Mizuki) of the HSC-SSP PDR3, across both Wide and
Deep/UltraDeep layers. Each PDZ is compressed via PCA and combined with summary descriptors; a multilayer perceptron, optimized with \texttt{Optuna} under a composite objective ($\sigma_{\rm NMAD}$ and $\eta_{0.15}$), produces the optimized point estimate $\zml$. A second network, trained in log-space with MAE and post-hoc calibrated, yields the empirical error-scale estimate $\sml$, from which the reliability score $\rml$ is derived via percentile ranking.
The optimized point estimate $\zml$ improves $\sigma_{\rm NMAD}$ and $\eta_{0.15}$ relative to the catalog $\zbest$ across all six pipeline--layer combinations. Absolute median bias is smaller in four configurations and tied in two. The reliability score $\rml$ gives the lowest retained-fraction AUC in 10 of the 12 native pipeline--layer/metric comparisons. For the Mizuki template-fitting pipeline, the catalog indicators perform particularly poorly, with AUC values up to about 13 times larger than those of $\rml$. Feature-importance analysis reveals complementary patterns: point estimation is dominated by the PCA block and location statistics, whereas reliability estimation gives greatest weight to peak, PCA, and shape information.
The pipeline is survey-agnostic by design, but each PDZ dataset requires its own fitted preprocessing, PCA basis, and trained models.
The code will be released as a reusable Python library at \href{https://github.com/valerio-marra/turboPDZ}{github.com/valerio-marra/turboPDZ}.
\end{abstract}


\section{Introduction}

The precise determination of photometric redshifts (\pz or $\zphot$) is crucial for mapping the large-scale structure (LSS) of the Universe and for the statistical analysis of galaxy surveys, whose ultimate goal is to constrain the nature of dark energy and dark matter, as well as the physics of the early Universe.
Photometric redshifts provide a cost-effective alternative to spectroscopic redshifts (\sz or $\zspec$), enabling the study of millions of galaxies over wide redshift ranges without the need for pre-selection; for a comprehensive review see
\citet{2019NatAs...3..212S}.
Modern surveys that rely extensively on \pz\ include DES \citep{DES:2025key}, HSC-SSP \citep{Aihara:2021jwb}, J-PAS \citep{Bonoli:2020ciz}, LSST \citep{2018arXiv180901669T}, and Euclid \citep{Euclid:2024yrr}.

Despite their central role, achieving both accuracy and reliability in \pz estimates remains challenging. State-of-the-art methods, based on either template fitting or machine learning, return for each galaxy a redshift probability distribution function (PDZ), which summarizes the inferred redshift information. However, LSS analyses often rely on point estimates extracted from the PDZs, combined with quality cuts designed to obtain purer, or \textit{gold}, subsamples. As a result, the full probabilistic information is not necessarily used directly.

This situation motivates the optimization of two key quantities: the point estimate $\zml$ and its associated reliability measure $\rml$. Our goal is therefore twofold. First, we optimize $\zml$ for scatter and outlier fraction. Second, we optimize $\rml$, based on the empirical error-scale estimate $\sml$, to filter galaxy samples by expected photo-$z$ quality. While previous works have investigated aspects of point estimates and quality metrics \citep{2015MNRAS.452.3710R, Tanaka:2017lit, 2020arXiv200301511N, LSSTDarkEnergyScience:2020nwm, DES:2020aks, Euclid:2020gbk, Hernan-Caballero:2021aig, Ferreira:2025mnb}, here we introduce a systematic optimization of these features using deep learning.

In this work, we focus on the Public Data Release 3 of the Hyper Suprime-Cam Subaru Strategic Program \citep[HSC-SSP;][]{Aihara:2017paw}. This survey combines depth and areal coverage with high-quality publicly available data, and provides three independent pipelines for producing PDZs, which allows us to test the robustness of our methodology. The release includes a million galaxies with PDZs and corresponding \sz, enabling the efficient training of deep neural networks (NN).

With upcoming Stage-IV surveys placing stringent requirements on photo-$z$ precision \citep{2022ARA&A..60..363N}, optimized point estimates and reliability metrics are increasingly critical.
We will release the pipeline as a reusable Python library at \href{https://github.com/valerio-marra/turboPDZ}{github.com/valerio-marra/turboPDZ}.
The framework is survey-agnostic by design and broadly applicable after retraining, but its preprocessing, PCA basis, hyperparameters, and network weights are fitted separately for each PDZ dataset. Future extensions could incorporate the optimized PDZ representation learned by the network into other downstream analyses.

This paper is organized as follows. Section~\ref{sec:data} presents the data used in this work, including the survey description, the photometric-redshift pipelines, and the spectroscopic reference sample. Section~\ref{sec:zml} describes the optimization of the point estimates, while Section~\ref{sec:rml} focuses on the optimization of the reliability measures. Section~\ref{sec:applicability} defines the color--magnitude applicability diagnostics intended for the value-added catalog. Section~\ref{sec:results} presents the results of our optimization, and Section~\ref{sec:conclusions} summarizes our findings and discusses future directions. Appendices~\ref{ap:common-residual}--\ref{ap:hp} provide the controlled common-residual comparison, paired-bootstrap assessment, and additional details on the hyperparameter optimization.

\section{Data} \label{sec:data}

\subsection{Survey description}

HSC-SSP is a three-tiered optical imaging survey with the 8.2m Subaru Telescope at Maunakea. Using the Hyper Suprime-Cam prime-focus imager (1.5 deg$^2$ field of view), it achieves sub-arcsecond image quality (median $i$-band seeing $\simeq 0.6$ arcsec), enabling precise shape and color measurements.

We use the third public data release (PDR3) \citep{Aihara:2021jwb}, which contains photometry for about 507 million objects in the Wide layer and 19 million in the Deep and UltraDeep (DUD) layers. PDR3 covers $\sim 1470$ deg$^2$ in total, of which 670 deg$^2$ reach full color and full depth in the five broad bands ($grizy$). In the Wide layer, the typical $5\sigma$ point-source depth is $i \sim 26$ mag, with total exposure times of 10–20 min per band. The Deep layer comprises four $\sim 7$ deg$^2$ fields imaged in $grizy$ plus narrow bands, reaching $i \sim 27$ mag with 1–3 h per band. The UltraDeep layer consists of COSMOS and SXDS, observed for 5–10 h per band with additional narrow bands, yielding depths $\sim 0.8$~mag deeper than Deep. This tiered design provides a unique combination of depth and area among current ground-based surveys.

The homogeneous calibration and extensive spectroscopic overlap make PDR3 an ideal dataset for testing and validating photometric-redshift methodologies.

\subsection{Photometric redshift pipelines}

\begin{figure}
\centering 
\includegraphics[trim={0 0 0 0}, clip, width= \linewidth]{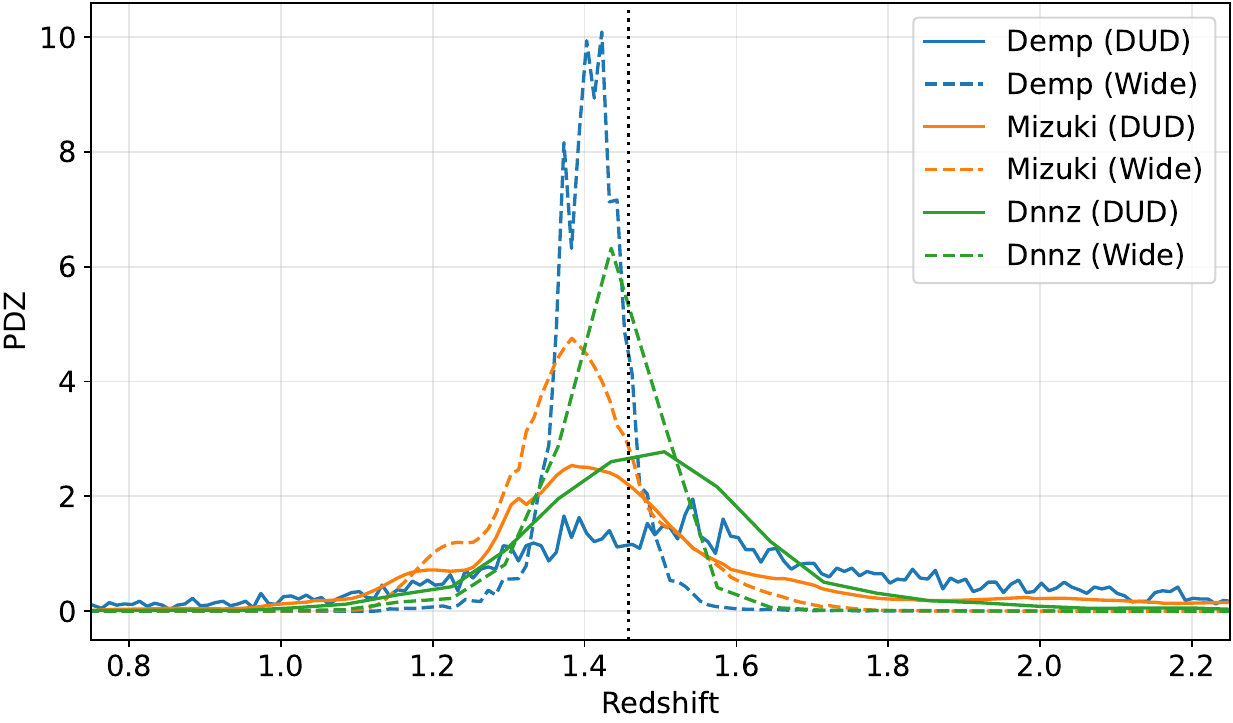}
\caption{
Example of the six PDZs available for a single galaxy, obtained from the three pipelines (DEmP, Mizuki, DNNz) in both the Wide and Deep/UltraDeep (DUD) layers. The comparison highlights systematic differences in shape, width, and modality across the pipelines. \\
{Alt text: Color line graph. Curves with varying shapes and peaks.}
\label{fig:ex}}
\end{figure}

The HSC-SSP PDR3 provides three independent photometric-redshift pipelines, each delivering full PDZs:
\begin{itemize}

\item \textbf{DEmP} \citep[Direct Empirical Photometric;][]{2014ApJ...792..102H} is a localized, data-driven regression. For each object, it identifies the 40 nearest training-set neighbors in a ten-dimensional space (five magnitudes, four colors, and an SDSS-like size parameter) and fits a local linear model to estimate the redshift. PDZs are generated by combining Monte Carlo perturbations of the photometry (500 realizations) with bootstrap resampling of the training set (500 per realization), and are sampled on 601 bins over $0 \leq z \leq 6$. Outputs are provided for all sources.

\item \textbf{DNNz} \citep{Nishizawa:pdr3} is a deep-learning method based on a multi-layer perceptron with five hidden layers. The input vector comprises 20 attributes per object: cModel fluxes, undeblended convolved fluxes, PSF fluxes, and a size estimate, across the five $grizy$ bands. Fluxes are converted to asinh magnitudes with appropriate transmission corrections. PDZs are produced on 100 bins over $0 \leq z \leq 7$ for all sources.

\item \textbf{Mizuki} \citep{2015ApJ...801...20T} is a Bayesian template-fitting code based on stellar population synthesis models. It applies redshift-dependent priors on galaxy properties and also estimates physical parameters such as stellar mass and star-formation rate. Fluxes are measured from undeblended, PSF-matched images and scaled to cModel totals, ensuring consistent colors. Photo-$z$ estimates are available for objects with $i<25$ in the Wide layer (about 210 million sources) and for almost all objects in the Deep and UltraDeep layers. PDZs are sampled on 701 bins spanning $0 \leq z \leq 7$.

\end{itemize}

The three pipelines each provide independent sets of PDZs for the Wide and DUD layers. Owing to the different observing conditions, we analyze the two layers separately. In total, this yields six distinct PDZ sets, which we use to test the robustness of our pipeline. As an illustration, Figure~\ref{fig:ex} shows the six PDZs obtained for the same object.

\subsection{Spectroscopic reference sample}

\begin{figure*}
\centering 
\includegraphics[trim={0 0 0 0}, clip, width= .49\linewidth]{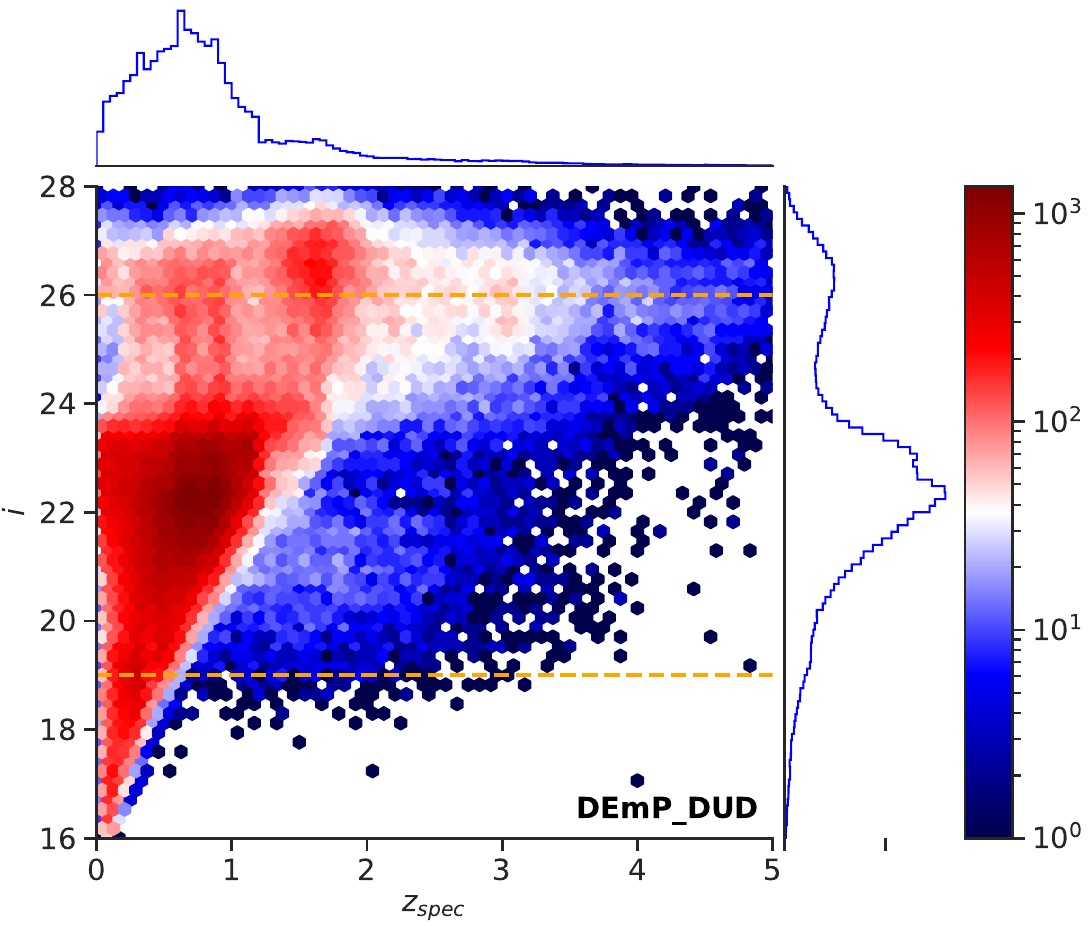}
\includegraphics[trim={0 0 0 0}, clip, width= .49\linewidth]{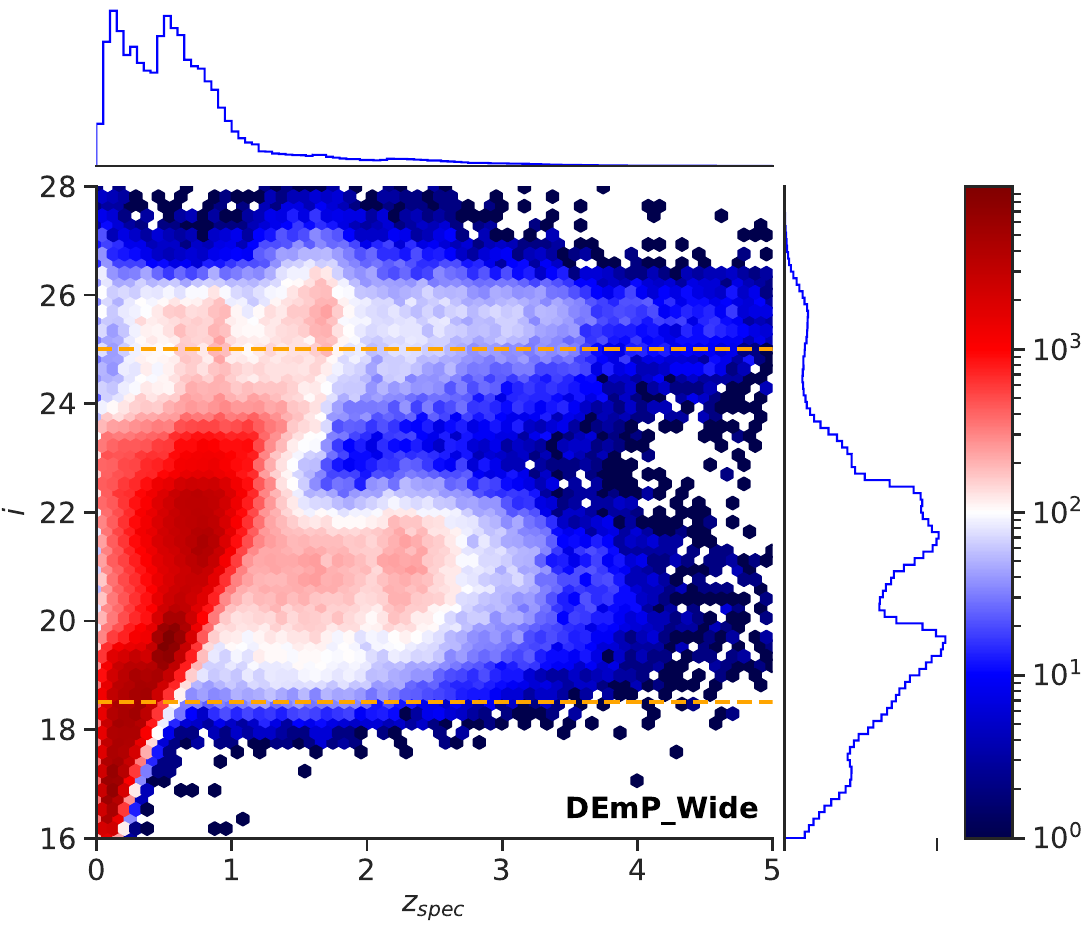}
\caption{
Spectroscopic redshift versus $i$-band magnitude for the labeled sets used in this work. The panels show the distributions for the DEmP pipeline in the DUD and Wide layers. The dashed lines indicate the applied cuts. The corresponding plots for Mizuki and DNNz are similar. \\
{Alt text: Scatter plot with two panels. Points show galaxies distributed across redshift and magnitude.}
\label{fig:zi-data}}
\end{figure*}

Photometric-redshift estimation in PDR3 is anchored to an extensive compilation of spectroscopic redshifts drawn from numerous public surveys overlapping the HSC-SSP footprint. The reference sample spans the full depth of the Wide and Deep/UltraDeep layers, with redshifts ranging from $z \simeq 0$ up to $z \gtrsim 6$, although the bulk of the sources lie at $z < 1.5$, as shown by Figure~\ref{fig:zi-data}.

We construct our labeled sets for training, validation, and testing by cross-matching via SQL the available PDZs\footnote{\url{https://hsc-release.mtk.nao.ac.jp/doc/index.php/photometric-redshifts__pdr3/}} with the relevant entries in the forced photometry, \texttt{photoz}, and \texttt{specz} tables available through the schema browser.\footnote{\url{https://hsc-release.mtk.nao.ac.jp/schema/}}

We require \texttt{specz\_flag\_homogeneous=True}, which identifies secure spectroscopic redshifts using HSC's harmonized survey-specific quality criteria. We use the released \texttt{specz} records and rely on HSC's duplicate resolution, which selects the measurement with the smallest quoted error when several secure redshifts are available.\footnote{\url{https://hsc-release.mtk.nao.ac.jp/doc/index.php/catalog-of-spectroscopic-redshifts__pdr3/}} Further details are given in the release documentation and papers \citep{Aihara:2021jwb,Nishizawa:pdr3}. Beyond the redshift and magnitude cuts specified below, we impose no additional spectroscopic-error or source-classification cuts. In particular, QSOs and AGNs present in the compilation are not explicitly excluded.

Each labeled set includes, for every source, the PDZ, \texttt{specz\_redshift}, and \texttt{i\_cmodel\_mag}, together with the point estimates \texttt{photoz\_X} and the associated reliability parameters \texttt{photoz\_conf\_X} and \texttt{photoz\_risk\_X}, where \texttt{X} corresponds to \texttt{mean}, \texttt{mode}, \texttt{median}, or \texttt{best}. The sizes of the resulting samples are listed in Table~\ref{tab:cat_summary}.

\begin{table}
\renewcommand{\vale}[1]{#1}
\centering
\setlength{\tabcolsep}{7pt}
\renewcommand{\arraystretch}{1.2}
\caption{Labeled sets before and after the homogeneous-spectroscopy, redshift, and magnitude cuts.}
\label{tab:cat_summary}
\begin{tabular}{llrr}
\toprule
Pipeline & Layer & Initial objects & After cuts \\
\midrule
DEmP   & DUD  & 236\,333    & 138\,718 \\
DEmP   & Wide & 1\,013\,670 & 571\,500 \\
DNNz   & DUD  & 236\,333    & 138\,537 \\
DNNz   & Wide & 1\,013\,670 & 541\,023 \\
Mizuki & DUD  & 233\,850    & 138\,679 \\
Mizuki & Wide & 838\,212    & 556\,606 \\
\bottomrule
\end{tabular}
\end{table}

\subsection{Preprocessing}

After requiring the homogeneous spectroscopic flag, we apply layer-specific cuts in spectroscopic redshift and $i$-band cModel magnitude to align the training domain with regimes where photo-$z$ estimates are informative and well calibrated. Table~\ref{tab:cat_summary} reports the number of galaxies after all three selections.

Wide layer: as in \citet{Tanaka:2017lit}, we restrict the sample to $0.01\leq\zspec\leq5$ to exclude stars at very low redshift. We limit the upper end at $z=5$ to avoid edge problems with the PDZs that are tabulated to redshift 6--7. We further restrict the sample to $18.5\leq i\leq25$: the bright-end cut avoids saturation \citep{Aihara:2021jwb}, and the faint-end cut excludes objects with unreliable photo-$z$ performance \citep{Tanaka:2017lit,2020arXiv200301511N,Nishizawa:pdr3}.

Deep/UltraDeep layers: we again use $0.01\leq\zspec\leq5$ and adopt $19\leq i\leq26$, reflecting the brighter saturation threshold in long exposures \citep{Aihara:2021jwb} and the $\sim1$~mag greater depth relative to Wide.

\section{Redshift point-estimate optimization} \label{sec:zml}

\subsection{Baseline point estimates}

PDZs in principle encode the available probabilistic redshift information for each galaxy. However, many applications require a single-number point estimate $\zphot$. Standard approaches extract such estimates directly from the PDZ:
\begin{itemize}
\item the \emph{mean}, which minimizes squared-error loss and is sensitive to extended tails;
\item the \emph{median}, the 50th percentile of the cumulative distribution, which is more robust to outliers but may be biased in asymmetric distributions;
\item the \emph{mode}, the redshift at which the PDZ peaks, which captures the most probable solution but is unstable in multimodal cases;
\item the \emph{best} estimate, the redshift at which the risk function introduced by \citet{Tanaka:2017lit} is minimized.
\end{itemize}
These baseline point estimates serve as natural references, and in what follows we use the `best' estimates as benchmarks for optimizing machine-learning-based predictions.

\subsection{Machine learning approach}
\label{sec:point-ml}

We developed a pipeline with a flexible architecture, designed to accommodate the markedly different types of PDZs produced by the various \pz pipelines.
Figure~\ref{fig:ex} illustrates this diversity by showing PDZs from different methods for the same galaxy.
The final configuration depends on the outcome of hyperparameter optimization, which is performed separately for each dataset under consideration using \texttt{Optuna} \citep{2019arXiv190710902A}.

The pipeline for the point-estimate optimization is:
\begin{enumerate}

\item  \textbf{PDZ compression.}
Since the PDZs are sampled with a variable number of bins and over different redshift domains, the pipeline begins with a dimensionality reduction step based on principal component analysis (PCA). This suppresses artificial structure introduced by oversampling, which otherwise produces highly correlated bins.
Also, this PDZ compression makes the optimization process and training faster.
The retained explained-variance (EV) fraction is optimized as a hyperparameter; the corresponding number $n$ of components is determined by fitting PCA on the relevant training pool.

\item \textbf{Feature construction.}
The model input combines:
\begin{itemize}
\item $n$ PCA components of the PDZ;
\item $i$-band cModel magnitude;
\item PDZ descriptors:
\begin{itemize}
\item central tendency: mean, median, mode;\footnote{We do not use the `best' descriptors for training, as they are generally unavailable in arbitrary catalogs.}
\item global shape: standard deviation, skewness, kurtosis, describing the overall width and asymmetry of the distribution;
\item peak structure: main-peak height, ratio of secondary to main peak height, distance of secondary peak to mode, peak influence (the product of the latter two), and second-peak flag,\footnote{The distance to mode, the peak influence, and the second-peak flag are all set to zero for unimodal PDZs.} which together quantify unimodality versus multimodality;
\end{itemize}
\end{itemize}

\item \textbf{Neural regressor.}
The feature vector 
is passed to a multilayer perceptron:
\begin{itemize}
\item dense feed-forward network; the number of hidden layers and the number of units per layer are optimized;
\item hidden-layer activation optimized between ReLU, ELU and Swish;
\item dropout after each hidden layer, with dropout rate  optimized;
\item \texttt{AdamW} optimizer (optimized batch size, learning rate and weight decay);
\item loss function: MAE;
\item output head: sigmoid neuron rescaled to the PDZ redshift upper bound, enforcing physically valid predictions in the sampled redshift interval.
\end{itemize}

\end{enumerate}

\subsection{Optimization and training}

{For each catalog/layer combination, we construct an 85/15 partition stratified in $\zspec$. The 15\% holdout is excluded from fitting the preprocessing and network weights of the $\zml$ model. The same partition is used for the with- and without-magnitude models and for the reliability analysis in Section~\ref{sec:rml}; for the latter, the holdout is divided equally into 7.5\% calibration and 7.5\% final-test subsets.
The network hyperparameters and epoch budgets are reused from previous searches on the spectroscopic sample before the secure-redshift cut. Those searches used \texttt{Optuna} with five-fold stratified cross-validation within their respective training pools, early stopping, learning-rate reduction, and pruning. Appendix~\ref{ap:hp} specifies the search samples and splits.}

The pipeline adopts a composite objective, defined as
\begin{equation} \label{Jcomp}
\mathcal{J}_{\rm comp}
=
\frac{1}{2}\,\frac{\sigma_{\rm NMAD}}{\sigma_{\rm NMAD, best}}
+
\frac{1}{2}\,\frac{\eta_{0.15}}{\eta_{\rm 0.15, best}} \,,
\end{equation}
where the reference values are computed from $\zbest$ on the training split, and the trial score is averaged over the five folds. We assign equal weights to scatter and outlier fraction, treating them as equally important for downstream scientific analyses. A separate bias term is omitted; consequently, the optimization does not guarantee a reduction in median residual, which is assessed independently on the held-out sample.

 {For the final model, feature scaling and PCA are fitted on the full training pool, and the selected network is trained on that pool using the epoch budget determined from cross-validation. Performance is then evaluated on the held-out test sample.}
Further implementation details, search spaces, and optimization controls are provided in Appendix~\ref{ap:hp}.

We also tested optional redshift-dependent sample weighting, including inverse-frequency weights and enhanced weights for the underrepresented but scientifically important population at $1.2 \lesssim z \lesssim 2.0$ (Figure~\ref{fig:zi-data}). No robust gain was found. Therefore, the reported baseline uses uniform weights, while weighting remains an option in the implementation.

\subsection{Evaluation framework}

Performance is evaluated against spectroscopic redshifts using standard metrics of accuracy, precision, and robustness.
Let $\Delta z$ denote the normalized residual,
\begin{equation}
\Delta z = \frac{\zphot - \zspec}{1 + \zspec} \, ,
\end{equation}
so that bias, scatter, outlier rate, and inlier rate are defined as
\begin{align}
b_z &= \mathrm{median}(\Delta z) \,, \\
\sigma_{\mathrm{NMAD}} &= 1.48 \, \mathrm{median}\!\left( \, \left| \Delta z - \mathrm{median}(\Delta z) \right| \, \right) \,, \\
\eta_{X} &= \frac{N\!\left(|\Delta z| > X\right)}{N_{\mathrm{tot}}} \,, \\
f_{Y} &= \frac{N\!\left(|\Delta z| < Y\right)}{N_{\mathrm{tot}}} \, ,
\end{align}
where $N_{\mathrm{tot}}$ is the total number of sources.  We adopt $X=0.15$ and $Y=0.02$.

These metrics provide complementary information: the bias measures systematic offsets, $\sigma_{\mathrm{NMAD}}$ quantifies the robust scatter around the median relation, $\eta_{0.15}$ gives the fraction of catastrophic outliers, and $f_{0.02}$ characterizes the inlier rate at a scale representative of the Wide-layer performance.

\section{Reliability measure optimization} \label{sec:rml}

\subsection{Definition and motivation}

Photometric-redshift pipelines are typically required to provide a reliability indicator or usability flag for the point estimates $\zphot$ \citep{Euclid:2020gbk}. Such metrics allow the construction of high-quality subsamples by trading completeness for purity, yielding a calibrated gold sample.

Common reliability measures include uncertainty estimates derived directly from the PDZ and pipeline-specific quality parameters. For instance, \mytt{photoz_std_best} provides the second moment of the PDZ around \mytt{photoz_best}, while the risk parameter \mytt{photoz_risk_best} corresponds to the minimum of the risk function defined in \citet{Tanaka:2017lit}. Another widely used metric is the confidence parameter \mytt{photoz_conf_best},
which is based on the odds statistic \citep{2000ApJ...536..571B}. The odds parameter is commonly used as a proxy for the probability that the relative photo-$z$ error lies below a chosen threshold $X$:
\begin{equation}
| \Delta z |= \frac{| \zphot - \zspec|}{1 + \zspec} < X \,,
\end{equation}
Conventionally, it is defined as the integrated PDZ mass within a symmetric interval around the point estimate:
\begin{equation}
\text{odds}_X = \int_{\zphot - X (1+\zphot)}^{\zphot + X (1+\zphot)} \text{PDZ}(z)\, \dif z\,.
\end{equation}
The choice of $X$ depends on the problem at hand, and values in the range $0.02 < X < 0.06$ have been used \citep{2014A&A...562A..86J,Hernan-Caballero:2021aig,2024A&C....4900886T}.
This window is scaled by $1+\zphot$, whereas the residual is normalized by $1+\zspec$, so the integration interval only approximately matches the event $|\Delta z|<X$. Consequently, $1-\text{odds}_X$ is an approximate individual error probability only when the PDZ is calibrated and the interval mismatch is small. A sufficiently large representative subsample then approximately obeys
\begin{equation}
\eta_X \approx 1 - \langle \text{odds}_X \rangle \,,
\end{equation}
A miscalibrated PDZ or a shifted population need not satisfy this relation, so odds is a heuristic rather than a guaranteed proxy for the outlier rate.

Here, we aim to derive a reliability score directly from the data using machine learning.
The objective is to obtain a score $\rml$ that correlates more tightly with the \pz precision $\sigma_{\mathrm{NMAD}}$ and the outlier rate $\eta$ than standard heuristic metrics.

\subsection{Reliability score}

We construct the reliability score via an optimized data-driven predictor of the redshift-error scale associated with \(\zml\), using the same network family and input feature groups described in Section~\ref{sec:point-ml}.
Specifically, using the \(n\) PCA components, the \(i\)-band magnitude, and the PDZ descriptors as input features, we train a multilayer perceptron to predict an error scale for each galaxy. Rather than regressing directly on \(|\Delta z_{\rm ml}|\), we train the network to predict
\begin{equation}
t \equiv \log\!\left(|\Delta z_{\rm ml}| + \epsilon\right),
\end{equation}
Here \(\epsilon\) is a small positive constant fixed to the 0.1\% quantile of the out-of-fold \(|\Delta z_{\rm ml}|\) values in the reliability-model training pool, usually about $10^{-5}$--$10^{-4}$.
The network output \(\hat{t}\) is obtained by minimizing the MAE between prediction and target in this transformed space. Working in log-space mitigates the large dynamic range of \pz\ errors and stabilizes the optimization, preventing the loss from being dominated by a small number of catastrophic outliers.
It is important to stress that our aim is to obtain a reliability score, not a probabilistic uncertainty estimate.
Specifically, log-space MAE training does not define a probabilistic model for the redshift errors.\footnote{This could be achieved using a Gaussian negative log-likelihood loss on the signed residuals, although tests showed that this approach decreases the reliability score performance. A reliable characterization of the \pz uncertainty is required for forward-modeling applications, where redshift errors must be propagated into cosmological observables. We leave this optimization for later work.}
Instead, we focus on predicting a single error scale that can be used to rank galaxies by reliability.

 {The network hyperparameters were selected by an earlier \texttt{Optuna} search on the spectroscopic sample before the secure-redshift cut, minimizing validation MAE in log-space, and are held fixed for the cross-fitted-target training described below.}
Denoting the network prediction by \(\hat{t}\), the raw error-scale proxy is
\begin{equation}
\tilde{\sigma}_{\rm ml} = \exp(\hat{t}) - \epsilon .
\end{equation}
To prevent target leakage, we construct the reliability-model targets with the following data flow.
\begin{enumerate}
\item  {\textbf{Outer split.} We reuse the stratified 85/15 partition constructed on the secure-redshift sample for the $\zml$ fit.}

\item \textbf{Cross-fitted point estimates.} We divide the 85\% pool into five stratified folds. For each fold, feature scaling and PCA are fitted on the other four folds, a new \(\zml\) network is trained on those rows, and predictions are generated only for the held-out fold. Consequently, every galaxy in the 85\% pool receives an out-of-fold prediction from a model that did not see that galaxy during training. Each fold model is trained on \(4/5\) of the pool, corresponding to 68\% of the full labeled sample.

\item \textbf{Reliability-model training.} The five held-out predictions are assembled into one complete out-of-fold vector. For every object in the 85\% pool, the target is therefore \(t=\log(|\Delta z_{\rm ml}^{\rm OOF}|+\epsilon)\). The \(\sml\) preprocessing and network weights are then fitted on these cross-fitted targets.

\item \textbf{Calibration and testing.} We divide the 15\% \(\zml\) holdout equally, using a stratified split, into a 7.5\% calibration subset and a 7.5\% final test subset. For both subsets, residuals are computed using the final \(\zml\) model trained on the full 85\% pool. Only the calibration subset is used to select the multiplicative calibration map and to construct the empirical reference CDF that defines \(\rml\); the test subset is used only for the final evaluation of \(\sml\) and \(\rml\).
\end{enumerate}

 {The final-test subset is excluded from fitting preprocessing and network weights, selecting the calibration map, and constructing the reliability reference distribution. Some of these objects were used in the earlier hyperparameter searches; Appendix~\ref{ap:hp} describes this limitation and the evaluation restricted to objects absent from those searches.}

\begin{figure*}
\centering 
\includegraphics[trim={0 0 0 0}, clip, width= \linewidth]{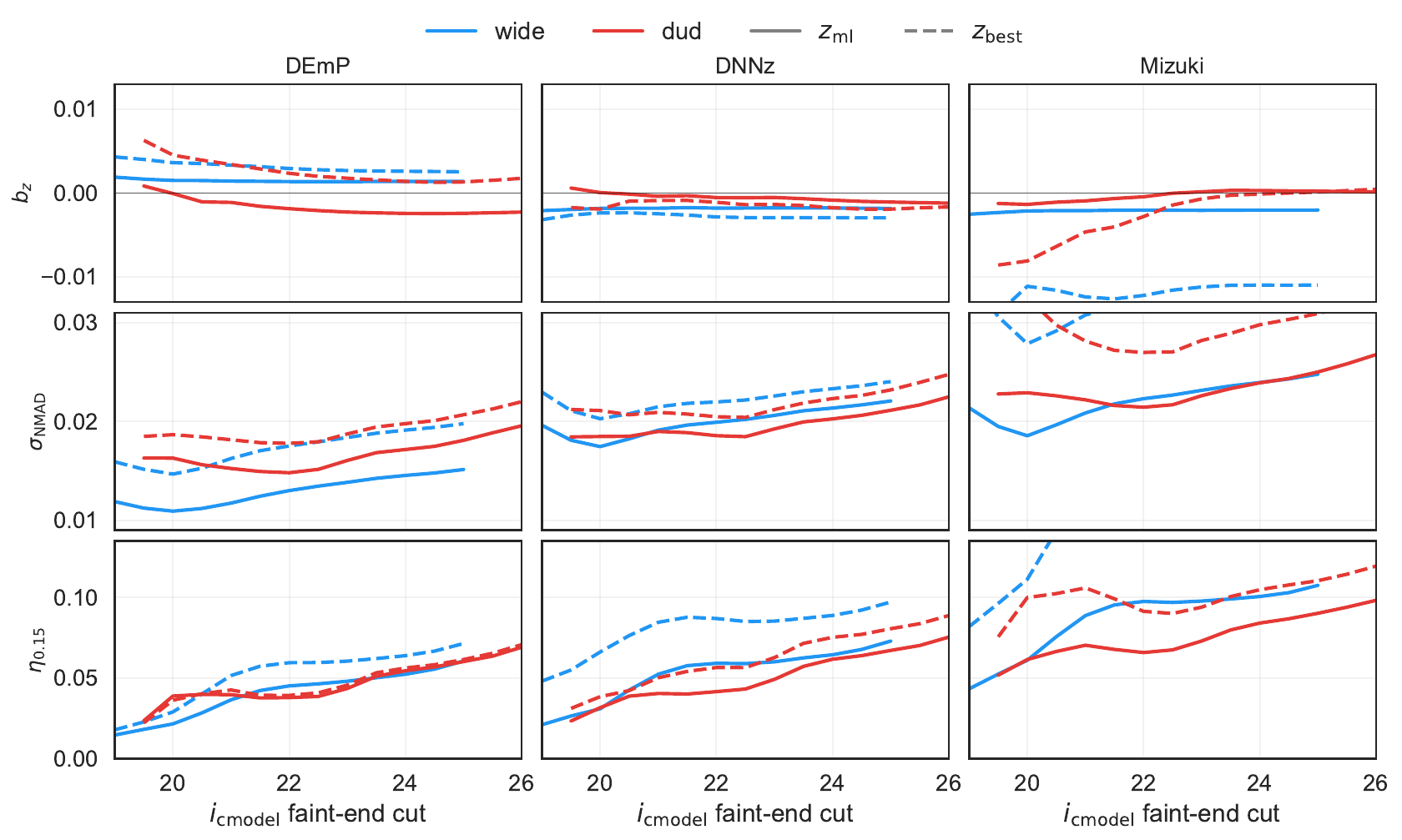}
\caption{
Cumulative \pz metrics as a function of the $i$-band magnitude cut; the $x$-axis denotes the faintest magnitude retained. \\
{Alt text: Multi-panel color line graph. Each panel shows a metric as a function of magnitude cut, with different colored lines for different pipelines/layers.}
\label{fig:metrics-cumulative-by-mag}}
\end{figure*}

To calibrate \(\tilde{\sigma}_{\rm ml}\) as an empirical predictor of the typical absolute residual, we use the  {separate} 7.5\% calibration subset.
We model a multiplicative correction
\begin{equation}
\sml = m(\tilde{\sigma}_{\rm ml})\,\tilde{\sigma}_{\rm ml},
\end{equation}
with $m(\cdot)$ estimated from calibration data only.
The resulting $\sml$ is a calibrated empirical error scale for the normalized residual $\Delta z_\mathrm{ml} = (\zml - \zspec )/(1+\zspec)$, not a posterior standard deviation or confidence interval.
In the ideal case, one would have \(m \simeq 1\) over the full range of \(\tilde{\sigma}_{\rm ml}\). In practice, deviations from unity quantify residual miscalibration of the raw network output. This calibration step\footnote{We do not apply an analogous calibration to \(\zml\), since its residual bias is already small (Figure~\ref{fig:metrics-cumulative-by-mag}).} therefore provides a data-driven rescaling that aligns the predicted calibrated error proxy with the empirically observed distribution of \(|\Delta z_{\rm ml}|\).

For a candidate calibration model, galaxies are sorted by $\tilde{\sigma}_{\rm ml}$ and partitioned into $n_{\rm bins}$ equal-count bins. In bin $j$, we define
\begin{equation}
m_j=\frac{\mathrm{median}\!\left(|\Delta z_{\rm ml}|\right)_j}
{\mathrm{median}\!\left(\tilde{\sigma}_{\rm ml}\right)_j}\,.
\end{equation}
The discrete sequence $\{m_j\}$ is then smoothed in $\log_{10}\tilde{\sigma}_{\rm ml}$ with a Gaussian kernel of bandwidth $w$ (in dex), yielding an interpolation table $(\sigma_{\rm anchor},m)$ used at inference time.

The calibration hyperparameters $n_{\rm bins}$ and $w$ are selected by four-fold cross-validation within the calibration sample, stratified in $\log_{10}\tilde{\sigma}_{\rm ml}$. Candidate maps are evaluated through $\langle |m_{\rm bin}-1|\rangle$ using 20 equal-count bins. The validation-score surface is lightly smoothed before selecting its minimum, subject to a tolerance that excludes solutions degraded by excessive smoothing. All calibration choices are made within this sample, leaving the final-test subset for evaluation.

\begin{table*}
\centering
\setlength{\tabcolsep}{2pt}
\renewcommand{\arraystretch}{1.2}
\caption{Selected hyperparameters for the $\zml$ model in each pipeline and survey layer.}
\label{tab:best_hp}
\begin{tabular}{lcccccccccc}
\toprule
Pipeline & Layer & Exp.~var. & PCA comps. & Batch size & \# Layers & Units/layer & Dropout & Learn.\ rate & Weight decay & Activation \\
\midrule
DEmP   & Wide & 0.79 &  {60} & 128 & 7 & 192--1024 & 0.11 & $4.3\times10^{-4}$ & $7.0\times10^{-4}$ & Swish \\
DEmP   & DUD  & 0.50 & 6  & 64  & 5 & 320--1024 & 0.11 & $4.6\times10^{-4}$ & $1.2\times10^{-4}$ & Swish \\
DNNz   & Wide & 0.79 &  {24} & 128 & 5 & 448--960  & 0.14 & $3.7\times10^{-4}$ & $1.4\times10^{-5}$ & Swish \\
DNNz   & DUD  & 0.77 &  {17} & 128 & 5 & 256--960  & 0.12 & $7.9\times10^{-4}$ & $6.7\times10^{-4}$ & Swish \\
Mizuki & Wide & 0.80 &  {22} & 128 & 6 & 512--1024 & 0.12 & $3.0\times10^{-4}$ & $6.2\times10^{-4}$ & Swish \\
Mizuki & DUD  & 0.80 &  {31} & 128 & 4 & 704--1024 & 0.12 & $7.6\times10^{-4}$ & $1.8\times10^{-6}$ & Swish \\
\bottomrule
\end{tabular}
\end{table*}

Finally, we convert the calibrated error estimate $\sml$ into a reliability score by ranking each galaxy relative to the calibration reference distribution. Let $F_{\sigma}$ denote the empirical cumulative distribution function of $\sml$ on the calibration split, and define
\begin{equation}
\rml = 1 - F_{\sigma}\!\left(\sml\right)\,.
\end{equation}
By construction, $\rml \in [0,1]$ is monotonic in the predicted error scale: smaller $\sml$ corresponds to larger $\rml$, so high-reliability objects are those assigned a smaller predicted error scale.

Because $\rml$ is defined from the empirical percentile rank, it is uniformly distributed on the calibration reference sample. This gives a direct mapping between a threshold cut and the retained fraction: a cut $\rml \ge t$ keeps approximately $1-t$ of the galaxies in the reference distribution. Equivalently, to retain a target fraction $f$, one may use the threshold $t \approx 1-f$. Here $\rml$ is a sample-dependent percentile rank, not an absolute probability that an individual redshift is correct; the retained-fraction identity is calibrated to the reference distribution and need not persist under domain shift.





\begin{table*}
\renewcommand{\vale}[1]{#1}
\centering
\setlength{\tabcolsep}{6pt}
\renewcommand{\arraystretch}{1.2}
\caption{Performance metrics for $\zml$ and $\zbest$ on the test set, for the full sample and restricted to $z<2$. Better displayed values are highlighted in bold red; ties are left unmarked.}
\label{tab:metrics}
\begin{tabular}{llcccccccccc}
\toprule
 &  & \multicolumn{5}{c}{$\zml$} & \multicolumn{5}{c}{$\zbest$} \\
\cmidrule(lr){3-7}\cmidrule(lr){8-12}
Pipeline & Layer & $b_z$ & $\sigma_{\rm NMAD}$ & $\eta_{0.15}$ & $f_{0.02}$ & MAE
              & $b_z$ & $\sigma_{\rm NMAD}$ & $\eta_{0.15}$ & $f_{0.02}$ & MAE \\
\midrule
\multicolumn{12}{l}{\textit{Full test set}} \\
\midrule
DEmP   & Wide & \textcolor{red}{\textbf{$\phantom{-}\bm{0.001}$}} & \textcolor{red}{\textbf{$\bm{0.015}$}} & \textcolor{red}{\textbf{$\bm{0.061}$}} & \textcolor{red}{\textbf{$\bm{0.690}$}} & \textcolor{red}{\textbf{$\bm{0.089}$}} & $\phantom{-}0.003$ & $0.020$ & $0.071$ & $0.627$ & $0.109$ \\
DEmP   & DUD  & $-0.002$ & \textcolor{red}{\textbf{$\bm{0.020}$}} & \textcolor{red}{\textbf{$\bm{0.069}$}} & \textcolor{red}{\textbf{$\bm{0.621}$}} & \textcolor{red}{\textbf{$\bm{0.106}$}} & $\phantom{-}0.002$ & $0.022$ & $0.071$ & $0.598$ & $0.118$ \\
DNNz   & Wide & \textcolor{red}{\textbf{$\bm{-0.002}$}} & \textcolor{red}{\textbf{$\bm{0.022}$}} & \textcolor{red}{\textbf{$\bm{0.073}$}} & \textcolor{red}{\textbf{$\bm{0.597}$}} & \textcolor{red}{\textbf{$\bm{0.107}$}} & $-0.003$ & $0.024$ & $0.097$ & $0.571$ & $0.131$ \\
DNNz   & DUD  & \textcolor{red}{\textbf{$\bm{-0.001}$}} & \textcolor{red}{\textbf{$\bm{0.022}$}} & \textcolor{red}{\textbf{$\bm{0.075}$}} & \textcolor{red}{\textbf{$\bm{0.588}$}} & \textcolor{red}{\textbf{$\bm{0.115}$}} & $-0.002$ & $0.025$ & $0.089$ & $0.559$ & $0.133$ \\
Mizuki & Wide & \textcolor{red}{\textbf{$\bm{-0.002}$}} & \textcolor{red}{\textbf{$\bm{0.025}$}} & \textcolor{red}{\textbf{$\bm{0.108}$}} & \textcolor{red}{\textbf{$\bm{0.552}$}} & \textcolor{red}{\textbf{$\bm{0.130}$}} & $-0.011$ & $0.034$ & $0.181$ & $0.454$ & $0.287$ \\
Mizuki & DUD  & $\phantom{-}0.000$ & \textcolor{red}{\textbf{$\bm{0.027}$}} & \textcolor{red}{\textbf{$\bm{0.098}$}} & \textcolor{red}{\textbf{$\bm{0.534}$}} & \textcolor{red}{\textbf{$\bm{0.137}$}} & $\phantom{-}0.000$ & $0.033$ & $0.120$ & $0.468$ & $0.172$ \\
\midrule
\multicolumn{12}{l}{\textit{Test set restricted to $z<2$}} \\
\midrule
DEmP   & Wide & \textcolor{red}{\textbf{$\phantom{-}\bm{0.002}$}} & \textcolor{red}{\textbf{$\bm{0.015}$}} & \textcolor{red}{\textbf{$\bm{0.051}$}} & \textcolor{red}{\textbf{$\bm{0.698}$}} & \textcolor{red}{\textbf{$\bm{0.068}$}} & $\phantom{-}0.003$ & $0.019$ & $0.055$ & $0.647$ & $0.079$ \\
DEmP   & DUD  & $-0.002$ & \textcolor{red}{\textbf{$\bm{0.018}$}} & \textcolor{red}{\textbf{$\bm{0.062}$}} & \textcolor{red}{\textbf{$\bm{0.641}$}} & \textcolor{red}{\textbf{$\bm{0.083}$}} & $\phantom{-}0.002$ & $0.021$ & $0.064$ & $0.618$ & $0.098$ \\
DNNz   & Wide & \textcolor{red}{\textbf{$\bm{-0.001}$}} & \textcolor{red}{\textbf{$\bm{0.021}$}} & \textcolor{red}{\textbf{$\bm{0.057}$}} & \textcolor{red}{\textbf{$\bm{0.615}$}} & \textcolor{red}{\textbf{$\bm{0.076}$}} & $-0.002$ & $0.023$ & $0.084$ & $0.587$ & $0.101$ \\
DNNz   & DUD  & \textcolor{red}{\textbf{$\bm{-0.001}$}} & \textcolor{red}{\textbf{$\bm{0.021}$}} & \textcolor{red}{\textbf{$\bm{0.068}$}} & \textcolor{red}{\textbf{$\bm{0.608}$}} & \textcolor{red}{\textbf{$\bm{0.091}$}} & $-0.002$ & $0.024$ & $0.084$ & $0.576$ & $0.114$ \\
Mizuki & Wide & \textcolor{red}{\textbf{$\bm{-0.001}$}} & \textcolor{red}{\textbf{$\bm{0.023}$}} & \textcolor{red}{\textbf{$\bm{0.082}$}} & \textcolor{red}{\textbf{$\bm{0.582}$}} & \textcolor{red}{\textbf{$\bm{0.091}$}} & $-0.009$ & $0.030$ & $0.134$ & $0.482$ & $0.162$ \\
Mizuki & DUD  & $\phantom{-}0.001$ & \textcolor{red}{\textbf{$\bm{0.025}$}} & \textcolor{red}{\textbf{$\bm{0.086}$}} & \textcolor{red}{\textbf{$\bm{0.558}$}} & \textcolor{red}{\textbf{$\bm{0.104}$}} & $\phantom{-}0.001$ & $0.031$ & $0.107$ & $0.485$ & $0.131$ \\
\bottomrule
\end{tabular}
\end{table*}

\subsection{Evaluation framework}

To assess the performance of $\sml$ and $\rml$, we compare them with the analogous quantities provided by the HSC-SSP pipelines following \citet{Tanaka:2017lit}. In particular, we compare $\sml$ with the PDZ-based error-scale proxies
\begin{align}
\sigma_{\rm best} &= \frac{\mytt{photoz_std_best}}{1+ \mytt{photoz_best}}  \,, \\
s_{\rm best} &= \frac{\mytt{photoz_err68_max} - \mytt{photoz_err68_min}}{2 (1+ \mytt{photoz_best})} \,, \label{sbest}
\end{align}
where \mytt{photoz_std_best} is the second-order moment around \mytt{photoz_best}, and \mytt{photoz_err68_min} and  \mytt{photoz_err68_max} are the 16th and 84th percentiles of the PDZ, respectively.
The normalization by $1+\zphot$ in these definitions reflects the fact that they are obtained, similarly to $\sml$, from the PDZ.

Since our goal is to rank galaxies by reliability rather than to obtain probabilistic uncertainty estimates, we evaluate $\sml$ through its ability to discriminate high- from low-error objects.
Specifically, we compute the Spearman and Pearson correlation coefficients between the realized absolute error $|\Delta z|$ and the predicted error-scale proxy $\sigma$; both should be positive for a useful estimator, with larger values indicating stronger discrimination power.
These metrics are reported for the full test set and in bins of $\zspec \in [0,\,0.5)$, $[0.5,\,0.8)$, and $\ge 0.8$. For $\sigma_{\rm best}$ and $s_{\rm best}$ we use $\Delta z_{\rm best}$, and for $\sml$ we use $\Delta z_{\rm ml}$.

Regarding the performance of $\rml$, we focus on the two key photo-$z$ quality metrics, $\sigma_{\mathrm{NMAD}}$ and $\eta_{0.15}$. Specifically, we test how efficiently $\rml$ performs relative to the pipeline-provided indicators \mytt{photoz_risk_best} and \mytt{photoz_conf_best} when used to rank and filter galaxies, aiming to achieve lower values of both metrics at fixed retained fraction.

To condense performance into a single number, we introduce the area under the $\sigma_{\rm NMAD}$ and $\eta_{0.15}$ vs.~retained-fraction curves:
\begin{align}
{\rm AUC}_\sigma &= 100 \int_{0}^{1} \sigma_{\mathrm{NMAD}}(f) \, \dif f \,,  \label{eq:auc_snmad}\\
{\rm AUC}_\eta &= 100 \int_{0}^{1} \eta_{0.15}(f) \, \dif f \,, \label{eq:auc_eta}
\end{align}
where $\sigma_{\mathrm{NMAD}}(f)$ and $\eta_{0.15}(f)$ are the scatter and outlier rate measured after keeping only the fraction $f$ of galaxies with the highest reliability scores (and analogously for the catalog indicators). Graphically, these areas correspond to the regions under the curves shown in Figure~\ref{fig:filter}.
Values are multiplied by $100$ for readability.
A perfect filter would drive both metrics to zero as soon as the worst objects are discarded, giving a small area; a filter uncorrelated with photo-$z$ quality yields a nearly flat curve and an area close to the no-rejection baseline. Smaller AUC therefore means that fewer galaxies must be rejected to reach a given improvement in precision, i.e., better filtering efficiency.

\section{Applicability domain and catalog flags} \label{sec:applicability}

A random split of the spectroscopic compilation measures interpolation performance under the selection function of that compilation; it does not by itself establish unrestricted transfer to the full HSC photometric population. We therefore construct a model-independent applicability reference from the objects satisfying \texttt{specz\_flag\_homogeneous=True} and the layer-specific redshift and magnitude cuts of Section~\ref{sec:data}. The reference uses only the observed five-band cModel photometry and is independent of the PDZs, the trained $\zml$ and $\sml$ networks, and their predictions. We use the DEmP joined table, which provides the broadest retained spectroscopic coverage, to construct one reference for Wide and one for DUD; the same layer-level reference can then accompany all three PDZ pipelines.

For every object with finite, non-null $grizy$ photometry, we form the four adjacent colors and robustly standardize them with respect to the corresponding spectroscopic reference $\mathcal S$, using the median and interquartile range (IQR) of each color in the reference sample. If a band's \texttt{\_isnull} flag is true, we treat its magnitude as missing regardless of the numerical value stored in the FITS column:
\begin{equation}
\boldsymbol c=(g-r,\,r-i,\,i-z,\,z-y), \qquad
\widetilde c_j=\frac{c_j-\operatorname{med}_{\mathcal S}(c_j)}{\operatorname{IQR}_{\mathcal S}(c_j)} .
\label{eq:app-colors}
\end{equation}
For a query galaxy with standardized color vector $\boldsymbol x$, let $d_{20}(\boldsymbol x)$ be its Euclidean distance to the 20th-nearest galaxy in the spectroscopic reference $\mathcal S$. Equivalently, it is the smallest radius centered on $\boldsymbol x$ that contains 20 reference galaxies. This radius is small where similar colors are common and large where they are rare. To establish what is typical of the reference, we calculate the same distance for each reference galaxy, excluding that galaxy itself so that the radius contains 20 other galaxies. Let $F_{\mathcal S}$ be the empirical cumulative distribution of these leave-one-out distances. We define the support percentile as
\begin{equation}
p_{\rm support}(\boldsymbol x)
=100\,F_{\mathcal S}\!\left[d_{20}(\boldsymbol x)\right] .
\label{eq:app-percentile}
\end{equation}
This percentile gives the percentage of reference galaxies whose 20-neighbor distance is no larger than that of $\boldsymbol x$. Values near 100 therefore indicate that the query lies in a relatively sparse or unusual part of the observed color distribution. We set the support boundary at $T_{99}=F_{\mathcal S}^{-1}(0.99)$ and accept a query when $d_{20}(\boldsymbol x)\leq T_{99}$, meaning that at least 20 reference galaxies lie within distance $T_{99}$ of it.

Standardization rescales each color but does not remove correlations among them. A fixed-radius neighborhood is spherical in the standardized four-dimensional space (and ellipsoidal in the original colors), yet the acceptance rule does not measure distance from a single center such as the sample median. Instead, it identifies points with 20 nearby reference galaxies. The accepted region can therefore be irregular or disconnected, following the observed color locus.

The planned value-added catalog will contain the following four fields: \texttt{turbopdz\_in\_mag\_domain}, which tests the inclusive $i$-band interval ($18.5\leq i\leq25$ for Wide and $19\leq i\leq26$ for DUD); \texttt{turbopdz\_specz\_support\_percentile}, which stores $p_{\rm support}$; \texttt{turbopdz\_in\_specz\_support\_99}, which tests valid five-band photometry and $d_{20}\leq T_{99}$; and \texttt{turbopdz\_applicability\_flag}, which is true only when both the magnitude-domain and 99\% color-support conditions hold. Missing, null-flagged, or non-finite five-band photometry receives no support percentile and fails the latter two flags. Retaining the continuous percentile also permits users to impose a stricter threshold, such as 95\%, without refitting the reference.

These quantities diagnose interpolation versus extrapolation relative to the observed homogeneous spectroscopic color--magnitude domain. They do not demonstrate that the spectroscopy is population-complete, remove its selection biases, or certify the accuracy of an individual prediction. The full HSC catalog is not processed in the present work; the archive-scale value-added catalog will be produced after publication, by evaluating these model-independent flags alongside $\zml$, $\sml$, and $\rml$.

\section{Results} \label{sec:results}

\subsection{Impact of point-estimate optimization}

\begin{figure*}
\centering 
\includegraphics[trim={0 0 0 0}, clip, width= \linewidth]{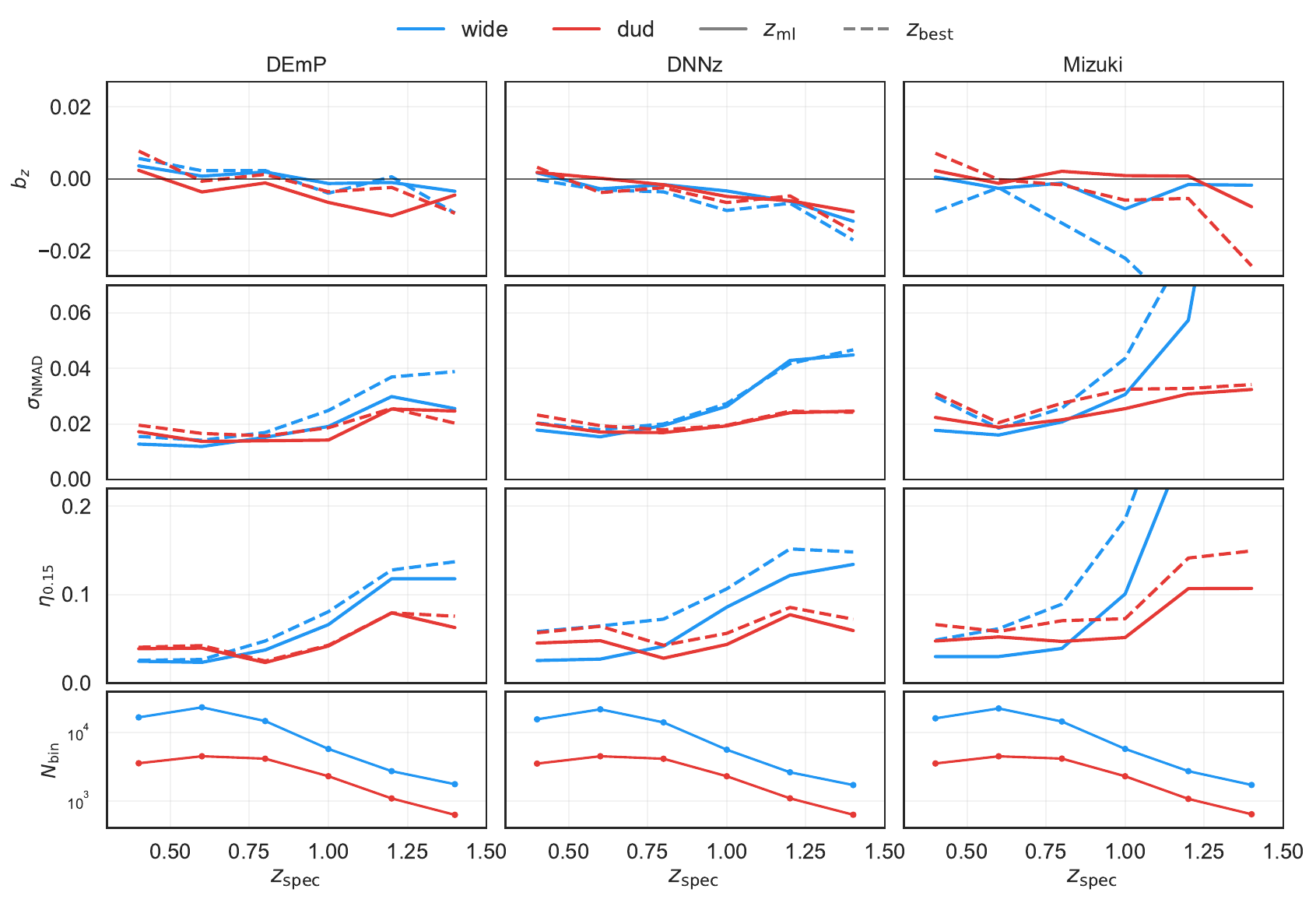}
\caption{
Photo-$z$ metrics versus \sz in non-cumulative bins of width 0.20. The bottom row shows the number of test galaxies in each redshift bin.\\
{Alt text: Line plots show median bias, scatter, outlier fraction, and redshift-bin counts for the held-out sample across three pipelines and two survey layers.}
\label{fig:metrics-by-specz}}
\end{figure*}

 {Table~\ref{tab:best_hp} summarizes the hyperparameters for each pipeline and survey layer. The network architectures are broadly similar across configurations, while the PDZ representations differ substantially in their PCA dimensionality. In the Wide layer, DNNz and Mizuki require only 24 and 22 components, respectively, to retain approximately 80\% of the variance, compared with 60 for DEmP, yielding considerably more compact representations.}
This architectural consistency---Swish activation and batch size of 64--128 across all configurations, with layer counts and unit ranges varying only modestly---is consistent with the searches locating a similar useful region of hyperparameter space, but does not by itself demonstrate convergence to a unique optimum.
Details on the hyperparameter optimization process are provided in Appendix~\ref{ap:hp}.

Table~\ref{tab:metrics} summarizes the overall performance of the optimized point estimate $\zml$ relative to the HSC-SSP catalog value $\zbest$:
$\zml$ improves MAE, $\sigma_{\rm NMAD}$, $\eta_{0.15}$, and $f_{0.02}$ in all six pipeline-layer configurations at the point-estimate level. The paired-bootstrap assessment in Appendix~\ref{ap:bootstrap} supports the MAE and $\sigma_{\rm NMAD}$ improvements in all six configurations and the $\eta_{0.15}$ improvement in five; the small DEmP-DUD outlier-fraction difference remains unresolved.
Because the composite objective contains no bias term, a smaller absolute median residual is not guaranteed. As reported in Table~\ref{tab:metrics}, $|b_z|$ is smaller for $\zml$ in four of six configurations and tied in DEmP-DUD and Mizuki-DUD, both on the full test sets and on the $z<2$ subsets.

This comparison is non-trivial because $\zbest$ is not an input: the combination of PDZ-derived features and $i$-band magnitude improves on the catalog point estimate.  {The no-magnitude ablation below tests whether these improvements persist using only PDZ-derived inputs.}
Figures~\ref{fig:metrics-cumulative-by-mag} and \ref{fig:metrics-by-specz} show the cumulative magnitude-cut and binned-redshift metrics on the held-out $\zml$ test subsets, respectively.
We focus on the redshift range $0.3 < z < 1.5$, which is used for weak-lensing analyses \citep{Dalal:2023olq}.
Across this range, $\zml$ generally performs better than $\zbest$ in the displayed scatter and outlier-fraction curves, although individual bins can differ.

To isolate the contribution of the only non-PDZ input, we repeated the point-estimate training after removing $i_{\rm cmodel}$.  {For each pipeline-layer pair, we reused the corresponding with-magnitude model's 85/15 pool/test indices on the secure-redshift sample, its selected architecture and hyperparameters, and final epoch budget, while refitting the feature scaler, PDZ scaler, PCA transformation, and neural-network weights from scratch.} On the common held-out test objects, removing $i_{\rm cmodel}$ increases the MAE by 0.2--9.1\%. The change in $\sigma_{\rm NMAD}$ ranges from $-3.6\%$ to $+7.2\%$, while $\eta_{0.15}$ increases by 0.5--9.5\%. The no-magnitude models still outperform $\zbest$ in MAE and $\sigma_{\rm NMAD}$ in all six configurations, and in $\eta_{0.15}$ in five. For DEmP-DUD the no-magnitude and catalog outlier fractions are tied.  {This ablation shows that improved point estimates can be obtained using PDZ-derived inputs alone, with an additional, configuration-dependent contribution from magnitude. The ablation measures the effect of removing magnitude while retaining the selected hyperparameters. Because magnitude and PDZ features contain overlapping information, their individual contributions cannot be uniquely separated by this comparison.}

The HSC PDR3 photo-$z$ documentation describes a five-fold division of its calibration sample \citep{Nishizawa:pdr3}. However, the public documentation does not establish whether the production PDZ associated with each galaxy in our spectroscopic sample was generated without using that galaxy during training or calibration, and we could not identify public object-to-fold provenance sufficient to reconstruct this. Any overlap could affect both $\zbest$ and the PDZ supplied to \mytt{turboPDZ}; consequently, its effect on their relative performance has no predetermined direction favoring either model.

\subsubsection*{Feature Importance and Interpretability Analysis}

\begin{figure}
\centering 
\includegraphics[trim={0 0 0 0}, clip, width= \linewidth]{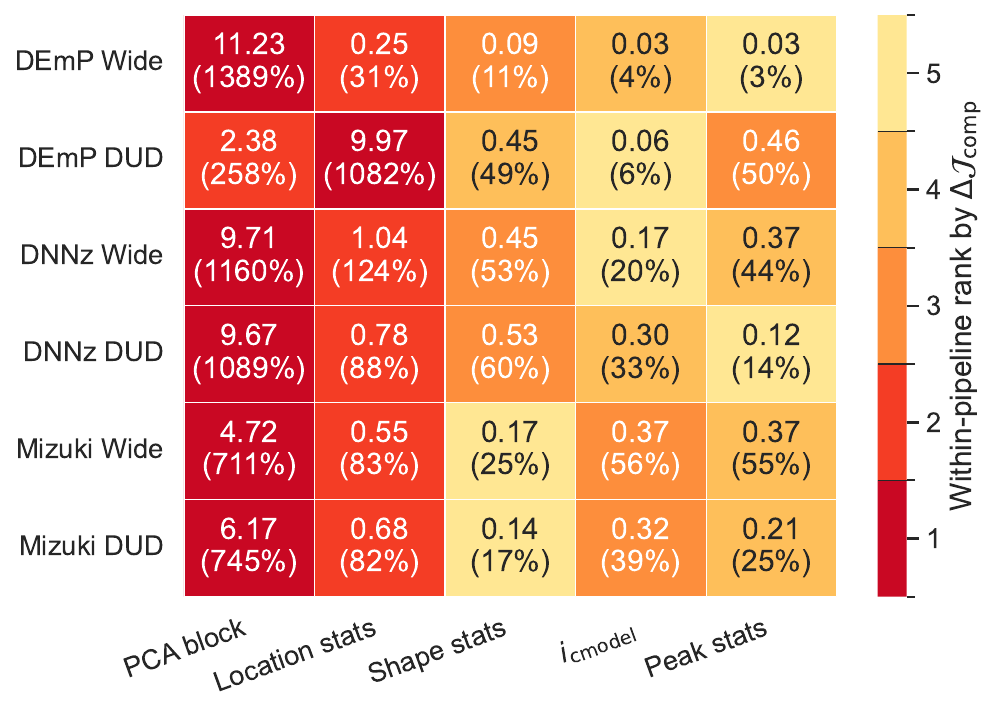}
\caption{
Grouped permutation feature importance for the $\zml$ model across the six pipeline-layer datasets, ordered by mean within-pipeline rank. Cell colors encode the within-pipeline rank class of each feature group according to $\Delta\mathcal{J}_{\rm comp}$ (rank 1 = highest importance). Cell annotations report both the raw $\Delta\mathcal{J}_{\rm comp}$ and its relative change (\%).\\
Alt text: Heatmap with six pipeline-layer rows and five feature-group columns. Cell colors show ranks, and annotations give the loss increase.
\label{fig:feature-importance-grouped-heatmap}}
\end{figure}

We quantify interpretability for the $\zml$ model with grouped and component-level permutation importance, using 8--12 repeats per test. Performance drops are measured relative to the unpermuted baseline in MAE and in $\mathrm{J}_{\rm comp}$ (Eq.~\ref{Jcomp}), with $\Delta\mathcal{J}_{\rm comp}$ as the primary ranking metric. Figure~\ref{fig:feature-importance-grouped-heatmap}
summarizes the grouped, cross-pipeline results across the six pipeline-layer datasets.

 {Across the six configurations, the PDZ-PCA block and location statistics occupy the top two permutation-importance ranks, averaging 1.2 and 1.8, respectively.}
The PDZ-PCA block produces the largest average increase in both objective and MAE, with location statistics second on average. This is physically plausible: mean/median/mode anchor the central redshift, while PCA adds complementary information on PDZ morphology (width, asymmetry, and multimodality). Shape statistics, $i_{\rm cmodel}$, and peak statistics follow with mean ranks of 3.8, 4.0, and 4.2, respectively; all remain informative but secondary.  {The relatively low permutation importance of $i_{\rm cmodel}$ may reduce sensitivity to differences between the labeled and full-catalog magnitude distributions, but it does not by itself establish robustness under such a shift.}

\begin{figure}
\centering 
\includegraphics[trim={0 0 29cm 0}, clip, width=.9 \linewidth]{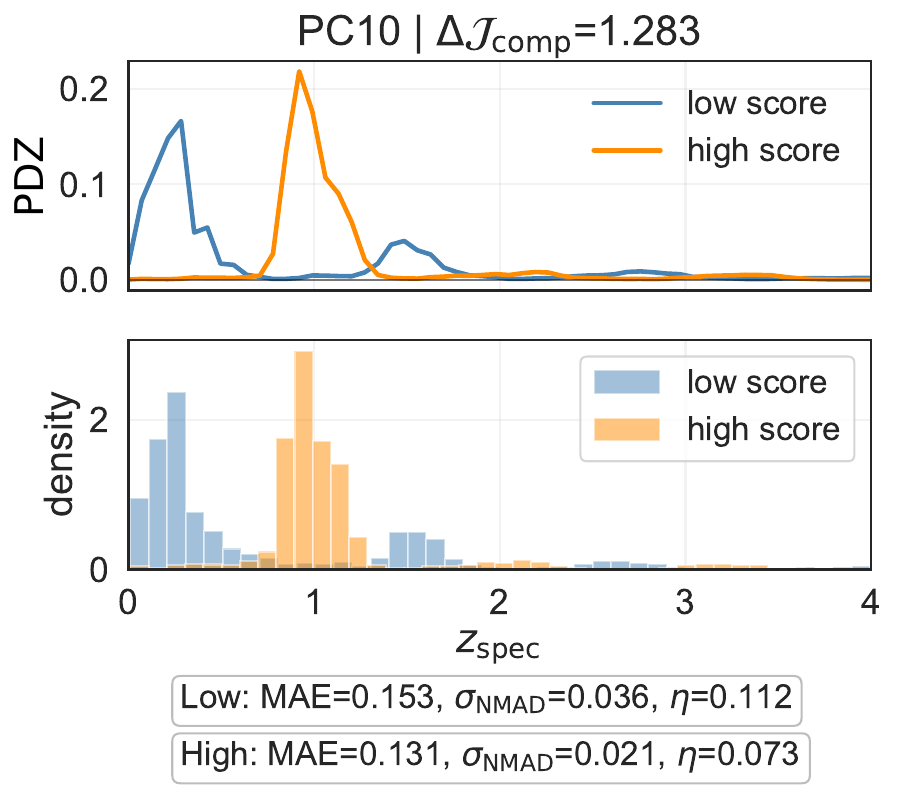}
\caption{
Representative DNNz-DUD slice analysis for PC10. Low and high scores denote the bottom and top 20\% of its score distribution. The upper panel compares their mean PDZs (blue and orange); the lower panel shows their $z_{\rm spec}$ distributions and photo-$z$ metrics. \\
Alt text: Two-panel figure. The upper panel compares the mean PDZ curves of low- and high-score objects. The lower panel compares their \sz histograms and lists MAE, $\sigma_{\rm NMAD}$, and outlier fraction.
\label{fig:pc_score}}
\end{figure}

To illustrate what the compressed PDZ representation retains, Figure~\ref{fig:pc_score} examines PC10 in the DNNz-DUD configuration, the component with the largest individual permutation impact. A galaxy's PC10 score is its coefficient along this PCA mode. We compare the bottom and top 20\% of scores in the held-out test set. “Low” and “high” refer only to the component's stored orientation: reversing its arbitrary PCA sign would exchange the labels without changing the separation or permutation impact. The upper panel compares the groups' mean PDZs; the lower panel shows their $z_{\rm spec}$ distributions and photo-$z$ metrics. Plots for other components, pipelines, and layers will accompany the public-library release in the \href{https://github.com/valerio-marra/turboPDZ}{turboPDZ repository}.

PC10 separates groups with distinct PDZ shapes and spectroscopic-redshift distributions, whose mean redshifts are $0.76$ and $1.16$ for the low- and high-score tails, respectively. The two peaks in the low-score group's mean PDZ predominantly reflect a mixture of low- and high-redshift galaxies, rather than two competing solutions in most individual PDZs. A small subset with intermediate $z_{\rm spec}$ receives little PDZ probability near its true redshift and contributes disproportionately to the low-score group's outliers. The high-score group consequently has lower MAE, $\sigma_{\rm NMAD}$, and $\eta_{0.15}$ despite its higher mean redshift. Together with the permutation result, this comparison shows that PC10 retains distinctions relevant to the fitted point estimator; it does not imply that PC10 score is a universal ranking of photo-$z$ quality.

\subsection{Impact of reliability measure optimization}

\begin{table*}
\centering
\setlength{\tabcolsep}{2pt}
\renewcommand{\arraystretch}{1.2}
\caption{Selected hyperparameters for the $\sml$ model in each pipeline and survey layer.}
\label{tab:best_hp_sigmaml}
\begin{tabular}{lcccccccccc}
\toprule
Pipeline & Layer & Exp.~var. & PCA comps. & Batch size & \# Layers & Units/layer & Dropout & Learn.\ rate & Weight decay & Activation \\
\midrule
DEmP   & Wide & 0.50 &  {12} & 512 & 3 & 192--512  & 0.12 & $2.2\times10^{-3}$ & {$1.9\times10^{-4}$} & Swish \\
DEmP   & DUD  & 0.68 & 12 & 128 & 3 & 320--960  & 0.11 & $1.4\times10^{-3}$ & {$1.3\times10^{-3}$} & ELU \\
DNNz   & Wide & 0.89 & 36 & 64  & 3 & 448--1024 & 0.30 & $1.1\times10^{-4}$ & $4.1\times10^{-6}$ & ReLU \\
DNNz   & DUD  & 0.90 &  {28} & 512 & 3 & 384--704  & 0.30 & $1.5\times10^{-3}$ & $1.3\times10^{-4}$ & ReLU \\
Mizuki & Wide & 0.90 &  {45} & 64  & 5 & 576--896  & 0.18 & $1.9\times10^{-4}$ & $5.7\times10^{-6}$ & ReLU \\
Mizuki & DUD  & 0.87 &  {46} & 256 & 3 & 576--1024 & 0.12 & $4.5\times10^{-4}$ & $4.4\times10^{-3}$ & ReLU \\
\bottomrule
\end{tabular}
\end{table*}

 {Table~\ref{tab:best_hp_sigmaml} reports the selected network hyperparameters for the $\sml$ analysis in each pipeline and survey layer.} A notable difference from the $\zml$ optimization is that the $\sml$ models are consistently shallower, with only 3--5 layers. This reflects the simpler, or perhaps noisier, nature of the error-scale prediction task compared to point estimation.

\begin{figure}
\centering 
\includegraphics[trim={0 0 0 0}, clip, width= \linewidth]{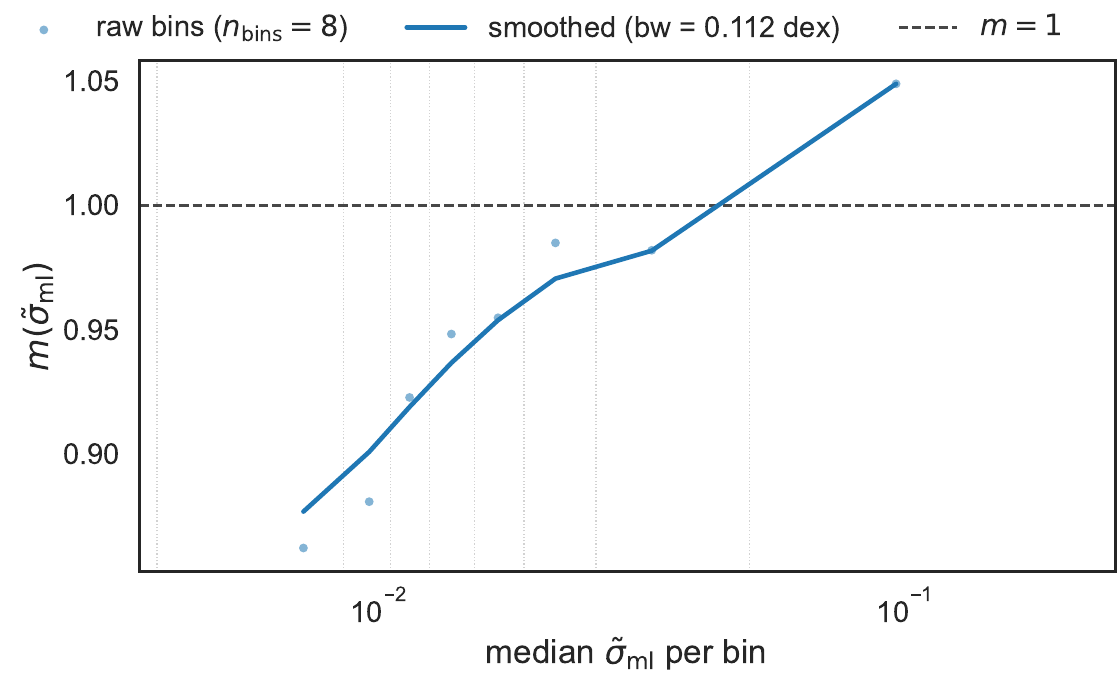}
\caption{
Calibration curve for the DNNz-Wide configuration. \\
{Alt text: Color line graph, close to unity across most of the horizontal axis.}
\label{calibration}}
\end{figure}

Figure~\ref{calibration} shows the calibration curve $m(\tilde{\sigma}_{\rm ml})$ for the DNNz-Wide configuration, with the raw network output $\tilde{\sigma}_{\rm ml}$ on the $x$-axis and the multiplicative correction $m$ on the $y$-axis. The curve is close to unity across most of the range, indicating modest corrections there, while deviations at low and high $\tilde{\sigma}_{\rm ml}$ are absorbed by the fitted multiplicative factor.

\begin{figure*}
\centering 
\includegraphics[trim={0 0 1.5cm 1.8cm}, clip, width= .49 \linewidth]{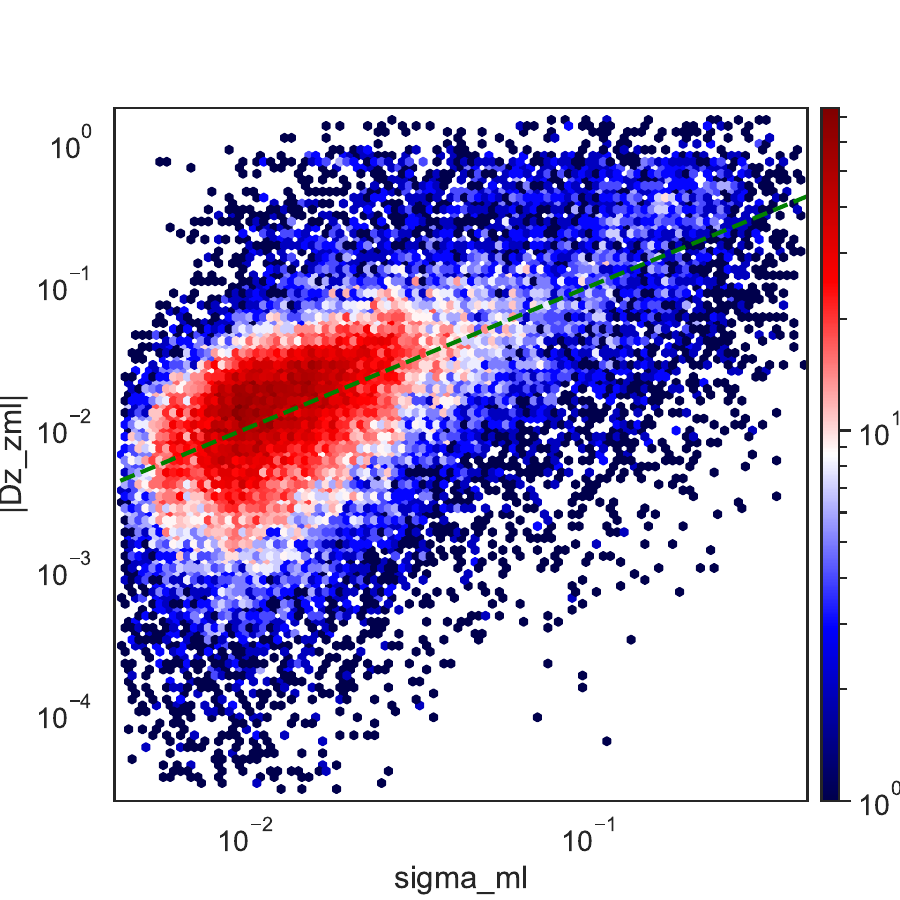}
\includegraphics[trim={0 0 1.5cm 1.8cm}, clip, width= .49 \linewidth]{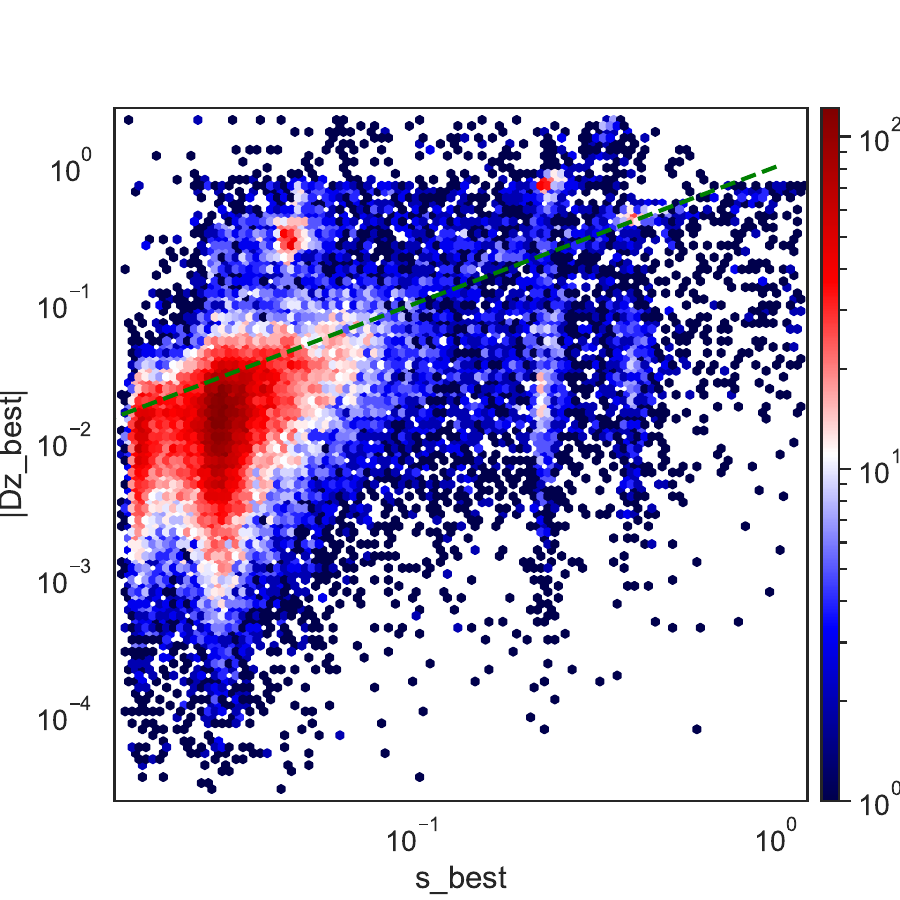}
\caption{
Comparison of the ML estimate $\sml$ with the catalog estimator $s_{\rm best}$ of Eq.~\eqref{sbest} on the test set for the DNNz-Wide configuration. \\
{Alt text: Two-panel color hexagonal binning density plot. Each panel shows predicted error scale versus absolute error, with color indicating point density.}
\label{sigma_ml_vs_s_best}}
\end{figure*}

Figure~\ref{sigma_ml_vs_s_best} compares the predicted error scale $\sml$ and the catalog estimator $s_{\rm best}$ of Eq.~\eqref{sbest} against the actual absolute error $|\Delta z|$ (we omit $\sigma_{\rm best}$ as it performs worse). The ML estimate yields a tighter, more uniform correlation with $|\Delta z|$, indicating stronger discrimination between high- and low-error objects. Table~\ref{tab:sigma_bins} quantifies this advantage via the Spearman and Pearson correlation coefficients between $|\Delta z|$ and the predicted error-scale proxy, evaluated across $\zspec$ bins for both estimators: $\sml$ has the larger coefficient in 10 of the 12 full-test Spearman/Pearson comparisons. The DEmP-Wide Pearson coefficients are tied at 0.42, while the DEmP-DUD Pearson coefficient favors $s_{\rm best}$ (0.28 versus 0.29). Paired-bootstrap intervals exclude zero in favor of $\sml$ for 8 of these 12 differences; the other four are unresolved, see Appendix~\ref{ap:bootstrap}. Within the three redshift bins, $\sml$ is larger in 30 of 36 comparisons. The exceptions are confined to DEmP and to the DNNz-DUD high-redshift Pearson coefficient.

\begin{table}
\renewcommand{\vale}[1]{#1}
\centering
\setlength{\tabcolsep}{2pt}
\renewcommand{\arraystretch}{1.0}
\caption{Error-scale estimator comparison in bins of $\zspec$, on the test set.
We report the Spearman ($\rho$) and Pearson ($r$) correlation coefficients between
$|\Delta z|$ and the predicted error scale. 
Bold red denotes the larger coefficient within each displayed pair; ties are left unmarked.
}
\label{tab:sigma_bins}
\begin{tabular}{lllrrrrr}
\toprule
Pipel. & Layer & $\zspec$ bin & $N$ & $\rho$($\sml$) & $\rho$($s_{\rm best}$) & $r$($\sml$) & $r$($s_{\rm best}$) \\
\midrule
DEmP & Wide & $[0,\,0.5)$    & 14\,238 & \better{0.61} & 0.57 & \better{0.43} & 0.37 \\
     &      & $[0.5,0.8)$  & 16\,038 & \better{0.56} & 0.52 & \better{0.48} & 0.44 \\
     &      & $\ge 0.8$      & 12\,587 & 0.57 & \better{0.58} & 0.39 & \better{0.51} \\
     &      & All            & 42\,863 & \better{0.59} & 0.58 & 0.42 & 0.42 \\
\cmidrule(lr){2-8}
     & DUD  & $[0,\,0.5)$    & 2\,987 & \better{0.54} & 0.51 & \better{0.31} & 0.25 \\
     &      & $[0.5,0.8)$  & 3\,341 & 0.49 & \better{0.51} & 0.22 & \better{0.36} \\
     &      & $\ge 0.8$      & 4\,076 & \better{0.62} & 0.58 & 0.35 & \better{0.52} \\
     &      & All            & 10\,404 & \better{0.57} & 0.55 & 0.28 & \better{0.29} \\
\midrule
DNNz & Wide & $[0,\,0.5)$    & 13\,490 & \better{0.52} & 0.45 & \better{0.45} & 0.27 \\
     &      & $[0.5,0.8)$  & 15\,027 & \better{0.46} & 0.38 & \better{0.44} & 0.33 \\
     &      & $\ge 0.8$      & 12\,060 & \better{0.58} & 0.49 & \better{0.41} & 0.39 \\
     &      & All            & 40\,577 & \better{0.56} & 0.49 & \better{0.43} & 0.34 \\
\cmidrule(lr){2-8}
     & DUD  & $[0,\,0.5)$    & 2\,972 & \better{0.50} & 0.47 & \better{0.31} & 0.23 \\
     &      & $[0.5,0.8)$  & 3\,340 & \better{0.46} & 0.43 & \better{0.49} & 0.22 \\
     &      & $\ge 0.8$      & 4\,079 & \better{0.53} & 0.48 & 0.44 & \better{0.48} \\
     &      & All            & 10\,391 & \better{0.52} & 0.48 & \better{0.31} & 0.27 \\
\midrule
Mizuki & Wide & $[0,\,0.5)$   & 13\,801 & \better{0.52} & 0.34 & \better{0.54} & 0.21 \\
       &      & $[0.5,0.8)$ & 15\,504 & \better{0.46} & 0.24 & \better{0.53} & 0.21 \\
       &      & $\ge 0.8$     & 12\,441 & \better{0.63} & $-0.56$ & \better{0.49} & $-0.09$ \\
       &      & All           & 41\,746 & \better{0.60} & $-0.02$ & \better{0.49} & 0.04 \\
\cmidrule(lr){2-8}
       & DUD  & $[0,\,0.5)$   & 2\,990 & \better{0.47} & 0.24 & \better{0.37} & 0.21 \\
       &      & $[0.5,0.8)$ & 3\,339 & \better{0.44} & 0.08 & \better{0.37} & 0.11 \\
       &      & $\ge 0.8$     & 4\,072 & \better{0.51} & 0.11 & \better{0.45} & 0.27 \\
       &      & All           & 10\,401 & \better{0.50} & 0.16 & \better{0.36} & 0.22 \\
\bottomrule
\end{tabular}
\end{table}

\begin{figure*}
\centering 
\includegraphics[trim={0 0 0 0}, clip, width= \linewidth]{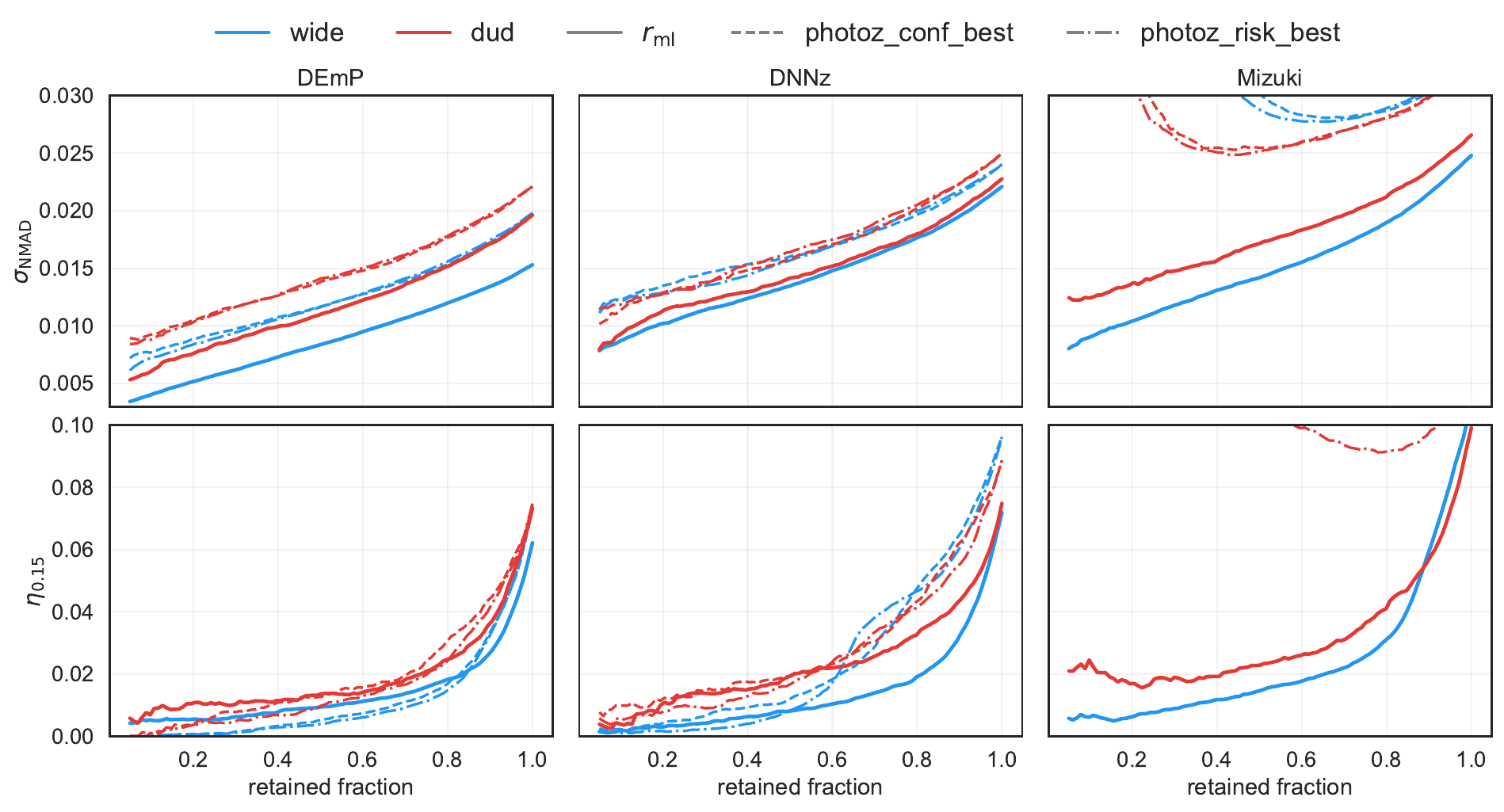}
\caption{
$\sigma_{\mathrm{NMAD}}$ and $\eta_{0.15}$ versus retained fraction filtering by $\rml$, \mytt{photoz_risk_best}, and \mytt{photoz_conf_best} across six pipeline-layer combinations.\\
{Alt text: Multi-panel color line graph. Each panel shows a metric as a function of retained fraction, with different colored curves for three filtering methods.}
\label{fig:filter}}
\end{figure*}

Figure~\ref{fig:filter} compares the performance of $\rml$ against the pipeline-provided indicators \mytt{photoz_risk_best} and \mytt{photoz_conf_best} as filters for constructing high-quality subsamples. We compute $\sigma_{\mathrm{NMAD}}$ and $\eta_{0.15}$ as functions of the retained fraction $f$, obtained by progressively tightening the threshold on each quantity. At $f=1$, all selectors retain the same test objects, but the native curves need not coincide: $\rml$ is evaluated using $\zml$ residuals, whereas the catalog indicators are evaluated using $\zbest$ residuals. As $f$ decreases, the curves generally diverge in favor of $\rml$. The integrated $\eta_{0.15}$ curves for DEmP-Wide and DEmP-DUD are the two exceptions, for which \mytt{photoz_risk_best} has the smaller area.

Table~\ref{tab:filter_auc} summarizes these findings via the metrics ${\rm AUC}_\sigma$ and ${\rm AUC}_\eta$, Eqs.~(\ref{eq:auc_snmad}-\ref{eq:auc_eta}), for each pipeline and survey layer. Smaller area indicates better filtering efficiency. We find that $\rml$ has the smallest area in 10 of the 12 pipeline-layer/metric comparisons. The exceptions are ${\rm AUC}_\eta$ for DEmP-Wide and DEmP-DUD. In the pairwise bootstrap of Appendix~\ref{ap:bootstrap}, all 12 native ${\rm AUC}_\sigma$ differences support $\rml$. For ${\rm AUC}_\eta$, 8 of 12 pairwise differences support $\rml$, 3 support the catalog selector, and the DEmP-DUD comparison with \mytt{photoz_conf_best} is unresolved.

Tables~\ref{tab:sigma_bins} and \ref{tab:filter_auc} report native end-to-end comparisons, in which the ML and catalog quantities accompany different point estimates. To separate the quality of those point estimates from the quality of the error-scale and reliability measures, Appendix~\ref{ap:common-residual} repeats the comparisons using, in turn, $|\Delta z_{\rm ml}|$ and $|\Delta z_{\rm best}|$ as a common residual. The main findings persist under both choices: $\sml$ has the higher correlation in 23 of 24 controlled coefficient comparisons, with 20 differences supported by paired-bootstrap intervals, and $\rml$ gives the lowest scatter-filtering AUC in all 12 controlled cases. The outlier-fraction filtering advantage is less uniform: $\rml$ gives the lowest AUC in 6 of 12 controlled cases. Thus, the stronger error association and scatter-oriented filtering are not solely consequences of improving the point estimate, whereas the outlier-filtering result remains pipeline dependent. Appendix~\ref{ap:bootstrap} reports the full paired intervals, including exceptions and unresolved differences.

\begin{table}
\renewcommand{\vale}[1]{#1}
\centering
\setlength{\tabcolsep}{5pt}
\renewcommand{\arraystretch}{1.05}
\caption{Area under the $\sigma_{\mathrm{NMAD}}$ (${\rm AUC}_\sigma$) and $\eta_{0.15}$ (${\rm AUC}_\eta$)
versus retained-fraction curves (see Fig.~\ref{fig:filter})
for selections based on $\rml$, \mytt{photoz_conf_best}, and \mytt{photoz_risk_best}.
Smaller area indicates better filtering efficiency. The best value per row and metric is shown in bold red.}
\label{tab:filter_auc}
\begin{tabular}{llrrrrrr}
\toprule
& & \multicolumn{3}{c}{${\rm AUC}_\sigma$} & \multicolumn{3}{c}{${\rm AUC}_\eta$} \\
\cmidrule(lr){3-5}\cmidrule(lr){6-8}
Pipeline & Layer & $\rml$ & conf & risk & $\rml$ & conf & risk \\
\midrule
DEmP   & Wide & \better{0.84} & 1.18 & 1.17 & 1.30 & 1.10 & \better{1.02} \\
DEmP   & DUD  & \better{1.11} & 1.37 & 1.37 & 1.78 & 1.79 & \better{1.55} \\
DNNz   & Wide & \better{1.35} & 1.60 & 1.58 & \better{1.29} & 2.39 & 2.25 \\
DNNz   & DUD  & \better{1.41} & 1.59 & 1.62 & \better{2.19} & 2.72 & 2.47 \\
Mizuki & Wide & \better{1.44} & 6.61 & 6.41 & \better{2.26} & 29.33 & 28.95 \\
Mizuki & DUD  & \better{1.70} & 3.65 & 3.51 & \better{2.99} & 15.49 & 14.42 \\
\bottomrule
\end{tabular}
\end{table}

The Mizuki pipeline presents a particularly instructive case. Here, the catalog indicators \mytt{photoz_risk_best} and \mytt{photoz_conf_best} perform poorly: $\sigma_{\mathrm{NMAD}}$ and $\eta_{0.15}$ can increase as the retained fraction decreases because these indicators preferentially retain some objects with worse photo-$z$ quality. The same behavior is visible in Table~\ref{tab:sigma_bins}, where $s_{\rm best}$ is anti-correlated with $|\Delta z|$ in the Mizuki-Wide high-redshift regime ($\rho=-0.56$) and remains weak in Mizuki-DUD.

As a diagnostic, we label an object suspicious when either $|\Delta z_{\rm best}|>0.2$ and \mytt{photoz_conf_best}$>0.9$, or \mytt{photoz_risk_best}$<0.005$. This selects 5\,643 of 41\,746 Mizuki-Wide test objects (13.5\%) and 784 of 10\,401 Mizuki-DUD objects (7.5\%). The selected sets contain 4\,954 of the 7\,534 Wide catastrophic outliers (65.8\%) and 477 of the 1\,232 DUD outliers (38.7\%). Their sharply concentrated but often incorrect PDZs are consistent with overconfident template solutions, although this association does not establish a unique failure mechanism. In contrast, $\rml$ uses the empirically learned relation between the full PDZ representation and realized error and ranks these cases more effectively.

\begin{figure}
\centering 
\includegraphics[trim={0 0 0 0}, clip, width= \linewidth]{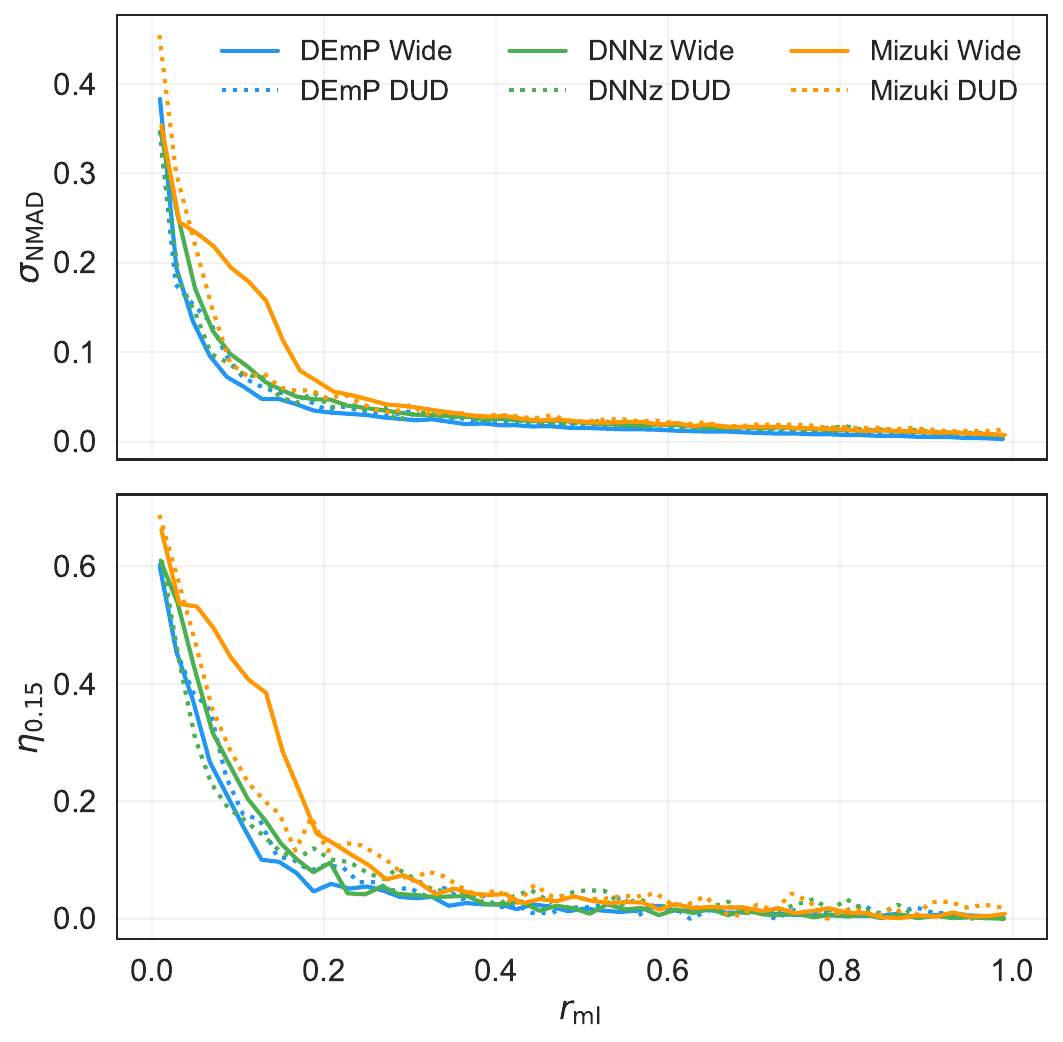}
\caption{
$\sigma_{\mathrm{NMAD}}$ and $\eta_{0.15}$ as functions of binned values of $\rml$. \\
{Alt text: Color line graph with two panels. Both metrics decrease as reliability score increases, with different colored lines for each pipeline/layer.}
\label{fig:rml}}
\end{figure}

Finally, Figure~\ref{fig:rml} shows $\sigma_{\mathrm{NMAD}}$ and $\eta_{0.15}$ as functions of binned $\rml$. The DEmP pipeline achieves the best overall performance, followed by DNNz and Mizuki. Both metrics show an overall decreasing trend with $\rml$, with local bin-to-bin fluctuations. This supports its use as a photo-$z$ quality ranking within the held-out spectroscopic sample.

\subsubsection*{Feature Importance}

\begin{figure}
\centering 
\includegraphics[trim={0 0 0 0}, clip, width= \linewidth]{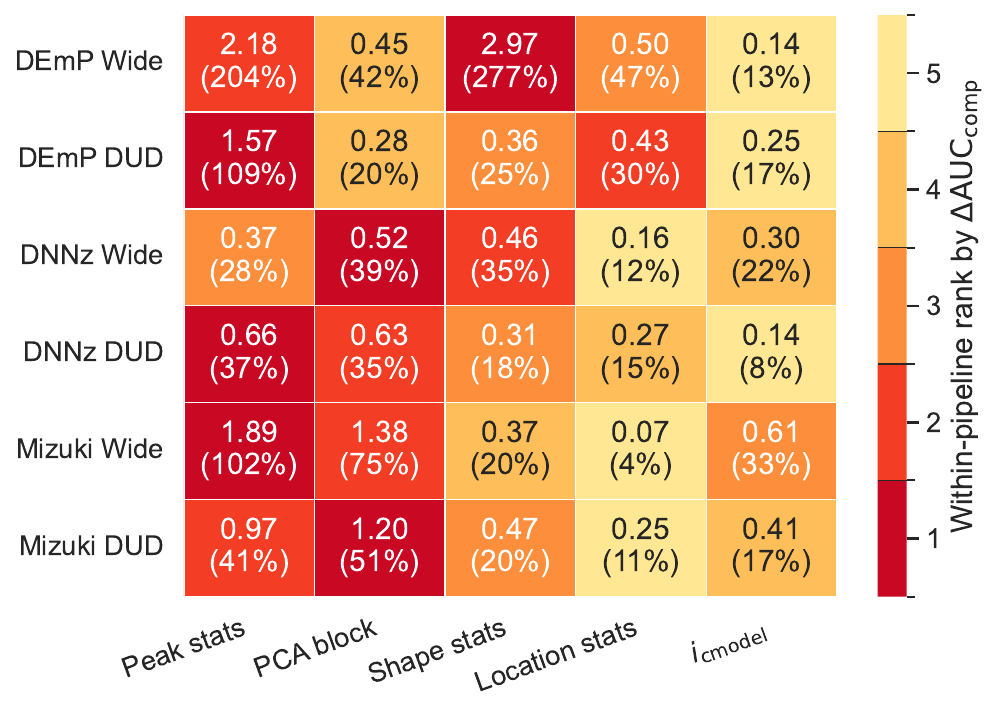}
\caption{
As in Figure~\ref{fig:feature-importance-grouped-heatmap} but for the $\rml$ model.
Alt text: as in Figure~\ref{fig:feature-importance-grouped-heatmap}.
\label{fig:feature-importance-grouped-heatmap-sigmaml}}
\end{figure}

We assess interpretability for the $\rml$ model with grouped permutation importance, using 8 repeats per test.  Performance degradation is measured relative to the unpermuted baseline through the increase $\Delta {\rm AUC}_{\rm comp}$ in the composite area ${\rm AUC}_{\rm comp} \equiv 0.5\,({\rm AUC}_\sigma + {\rm AUC}_\eta)$---see Table~\ref{tab:filter_auc} for the baseline values.  Figure~\ref{fig:feature-importance-grouped-heatmap-sigmaml} summarizes the cross-pipeline results across all six pipeline-layer configurations.

The feature-importance pattern for the $\rml$ model differs markedly from that of the point-estimate model $\zml$. For $\zml$, the PDZ-PCA block and location statistics dominate. For $\rml$, peak statistics rank first on average (1.7), followed by the PCA block (2.3) and shape statistics (2.7), while location statistics fall to 4.0. The $i$-band magnitude remains secondary (mean rank 4.3).
This shift is qualitatively interpretable. For point estimation, the central-tendency descriptors directly anchor the redshift prediction, whereas for error-scale prediction they contain less information about whether a PDZ is sharp, broad, or multimodal. Peak, shape, and PCA features encode this morphology and therefore become relatively more useful. Because these groups are correlated, however, permutation importance measures predictive dependence under the fitted feature representation and should not be read as a unique causal decomposition.

\section{Conclusions} \label{sec:conclusions}

We have presented \mytt{turboPDZ}, a machine-learning framework that optimizes the extraction of information from photometric-redshift probability distribution functions. Using PDZs from the three independent pipelines (DEmP, DNNz, Mizuki) of the HSC-SSP PDR3, across both the Wide and Deep/UltraDeep layers, we demonstrated consistent improvements in two complementary directions.

First, our optimized point estimate $\zml$ outperforms the catalog $\zbest$ in scatter, outlier fraction, and MAE across all six pipeline-layer combinations, despite not using $\zbest$ as an input feature.
Paired-bootstrap intervals support all six MAE and scatter improvements and five of the six outlier-fraction improvements; the small DEmP-DUD outlier-fraction difference is unresolved.
Retraining all six configurations without $i_{\rm cmodel}$ preserves the improvements in MAE and $\sigma_{\rm NMAD}$ over $\zbest$ in all six cases and in $\eta_{0.15}$ in five. This confirms that the gain is not driven solely by the additional magnitude input.
Second, our reliability score $\rml$, derived from the calibrated empirical error-scale estimate $\sml$, provides a more effective filter than the catalog indicators \mytt{photoz_risk_best} and \mytt{photoz_conf_best}. Quantitatively, $\rml$ has the smaller area under the $\sigma_{\mathrm{NMAD}}$ and $\eta_{0.15}$ versus retained-fraction curves in 10 of the 12 native pipeline-layer/metric comparisons; \mytt{photoz_risk_best} is better for ${\rm AUC}_\eta$ in DEmP-Wide and DEmP-DUD. The Mizuki pipeline is particularly revealing: the catalog-indicator AUC values reach about 13 times those of $\rml$ and become anti-correlated with photo-$z$ quality in the high-redshift regime, while $\rml$ retains a correctly oriented ranking.
The paired analysis supports $\rml$ in every native and controlled scatter-AUC comparison, whereas the outlier-AUC advantage remains pipeline dependent.

The feature-importance patterns of the two models reveal a qualitative complementarity. For point estimation, the dominant groups are the PCA block and location statistics. For reliability estimation, peak statistics rank first on average, followed by PCA and shape information, while location statistics and $i$-band magnitude are secondary. This is consistent with error scale depending strongly on PDZ morphology, but note that correlated-group permutation importance does not provide a causal decomposition.

The \mytt{turboPDZ} framework is survey-agnostic by design and can be applied to other PDZ datasets after fitting dataset-specific preprocessing, PCA bases, and models. The code will be released as a reusable Python library, together with the trained models and layer-specific applicability-reference artifacts, at \href{https://github.com/valerio-marra/turboPDZ}{github.com/valerio-marra/turboPDZ}. The archive-scale HSC-SSP PDR3 catalog containing $\zml$, $\sml$, $\rml$, and the applicability fields will be made available subsequently.
Future applications include extension to upcoming surveys, where reliable \pz point estimates and quality flags are essential for weak-lensing and large-scale-structure analyses.

\begin{ack}

We thank Sogo Mineo and Masayuki Tanaka for their assistance with accessing the HSC-SSP data, and Rodrigo von Marttens, Ribamar dos Reis, Wiliam Hipolito for useful discussions.\\
The authors would like to acknowledge the use of the computational resources provided by the \href{https://computacaocientifica.ufes.br/scicom}{Sci-Com Lab} of the Department of Physics at UFES, which was funded by FAPES, CAPES and CNPq.\\
The Hyper Suprime-Cam (HSC) collaboration includes the astronomical communities of Japan and Taiwan, and Princeton University. The HSC instrumentation and software were developed by the National Astronomical Observatory of Japan (NAOJ), the Kavli Institute for the Physics and Mathematics of the Universe (Kavli IPMU), the University of Tokyo, the High Energy Accelerator Research Organization (KEK), the Academia Sinica Institute for Astronomy and Astrophysics in Taiwan (ASIAA), and Princeton University. Funding was contributed by the FIRST program from the Japanese Cabinet Office, the Ministry of Education, Culture, Sports, Science and Technology (MEXT), the Japan Society for the Promotion of Science (JSPS), Japan Science and Technology Agency (JST), the Toray Science Foundation, NAOJ, Kavli IPMU, KEK, ASIAA, and Princeton University. \\
This paper makes use of software developed for Vera C. Rubin Observatory. We thank the Rubin Observatory for making their code available as free software at \href{http://pipelines.lsst.io/}{pipelines.lsst.io}.\\
This paper is based on data collected at the Subaru Telescope and retrieved from the HSC data archive system, which is operated by the Subaru Telescope and Astronomy Data Center (ADC) at NAOJ. Data analysis was in part carried out with the cooperation of Center for Computational Astrophysics (CfCA), NAOJ. We are honored and grateful for the opportunity of observing the Universe from Maunakea, which has the cultural, historical and natural significance in Hawaii.

\end{ack}

\section*{Funding}
RD thanks CAPES (Brazil) for financial support.
VM thanks CNPq (Brazil), CAPES (Brazil) and FAPES (Brazil) for partial financial support.

\section*{Data availability}
The HSC input data used in this article are publicly available.
The HSC-SSP PDR3 photometry, photometric redshifts, and spectroscopic redshifts were accessed via the HSC data archive system at \url{https://hsc-release.mtk.nao.ac.jp/}.
The code will be released as a reusable Python library, together with the trained models and layer-specific applicability-reference artifacts, at \href{https://github.com/valerio-marra/turboPDZ}{github.com/valerio-marra/turboPDZ}. The archive-scale HSC-SSP PDR3 catalog containing $\zml$, $\sml$, $\rml$, and the applicability fields will be made available subsequently.

\appendix

\section{Controlled common-residual comparison}
\label{ap:common-residual}

The native comparisons in Tables~\ref{tab:sigma_bins} and \ref{tab:filter_auc} quantify the performance of the complete point-estimate and reliability pipelines. However, they also combine two effects: the quality of the point estimate that defines the residual and the quality of the error-scale or ranking quantity. We therefore repeat both comparisons using a common residual target. For each pipeline-layer configuration, all quantities are restricted to the same joint finite subset of the final test sample and are evaluated first against $|\Delta z_{\rm ml}|$ and then against $|\Delta z_{\rm best}|$.

Table~\ref{tab:common-residual-correlations} compares $\sml$ and $s_{\rm best}$ with each common residual. The ML error-scale predictor has the larger coefficient in 23 of the 24 comparisons (two coefficients for each of 12 rows). The only exception is the Spearman coefficient for DEmP-Wide against $|\Delta z_{\rm best}|$, where $s_{\rm best}$ is slightly larger. The paired intervals in Table~\ref{tab:bootstrap-correlations} (Appendix~\ref{ap:bootstrap}) support $\sml$ in 20 comparisons, support $s_{\rm best}$ for that DEmP-Wide Spearman exception, and leave three DEmP Pearson differences unresolved. Thus, the stronger association of $\sml$ with realized error is not an artifact of pairing it only with $\zml$.

\begin{table}
\renewcommand{\vale}[1]{#1}
\centering
\footnotesize
\setlength{\tabcolsep}{2pt}
\renewcommand{\arraystretch}{1.05}
\caption{Controlled error-scale-estimator comparison. Spearman ($\rho$) and Pearson ($r$) coefficients are evaluated between each residual target and $\sml$ or $s_{\rm best}$. The best value is shown in bold red.}
\label{tab:common-residual-correlations}
\begin{tabular}{lllrrrrr}
\toprule
Pipeline & Layer & Residual & $N$ & $\rho$ ($\sml$) & $\rho$ ($s_{\rm best}$) & $r$ ($\sml$) & $r$ ($s_{\rm best}$) \\
\midrule
DEmP   & Wide & $|\Delta z_{\rm ml}|$   & 42\,863 & \better{0.59} & 0.42 & \better{0.42} & 0.33 \\
              &              & $|\Delta z_{\rm best}|$ & 42\,863 & 0.56 & \better{0.58} & \better{0.43} & 0.42 \\
\cmidrule(lr){2-8}
              & DUD  & $|\Delta z_{\rm ml}|$   & 10\,404 & \better{0.57} & 0.51 & \better{0.28} & 0.25 \\
              &              & $|\Delta z_{\rm best}|$ & 10\,404 & \better{0.58} & 0.55 & \better{0.34} & 0.29 \\
\midrule
DNNz   & Wide & $|\Delta z_{\rm ml}|$   & 40\,577 & \better{0.56} & 0.48 & \better{0.43} & 0.36 \\
              &              & $|\Delta z_{\rm best}|$ & 40\,577 & \better{0.56} & 0.49 & \better{0.49} & 0.34 \\
\cmidrule(lr){2-8}
              & DUD  & $|\Delta z_{\rm ml}|$   & 10\,391 & \better{0.52} & 0.47 & \better{0.31} & 0.24 \\
              &              & $|\Delta z_{\rm best}|$ & 10\,391 & \better{0.52} & 0.48 & \better{0.36} & 0.27 \\
\midrule
Mizuki & Wide & $|\Delta z_{\rm ml}|$   & 41\,746 & \better{0.60} & 0.01 & \better{0.49} & 0.15 \\
              &              & $|\Delta z_{\rm best}|$ & 41\,746 & \better{0.62} & $-0.02$ & \better{0.72} & 0.04 \\
\cmidrule(lr){2-8}
              & DUD  & $|\Delta z_{\rm ml}|$   & 10\,401 & \better{0.50} & 0.16 & \better{0.36} & 0.24 \\
              &              & $|\Delta z_{\rm best}|$ & 10\,401 & \better{0.47} & 0.16 & \better{0.38} & 0.22 \\
\bottomrule
\end{tabular}
\end{table}

\begin{figure*}
\centering 
\includegraphics[trim={0 1.35cm 0 .9cm}, clip, width=.9 \linewidth]{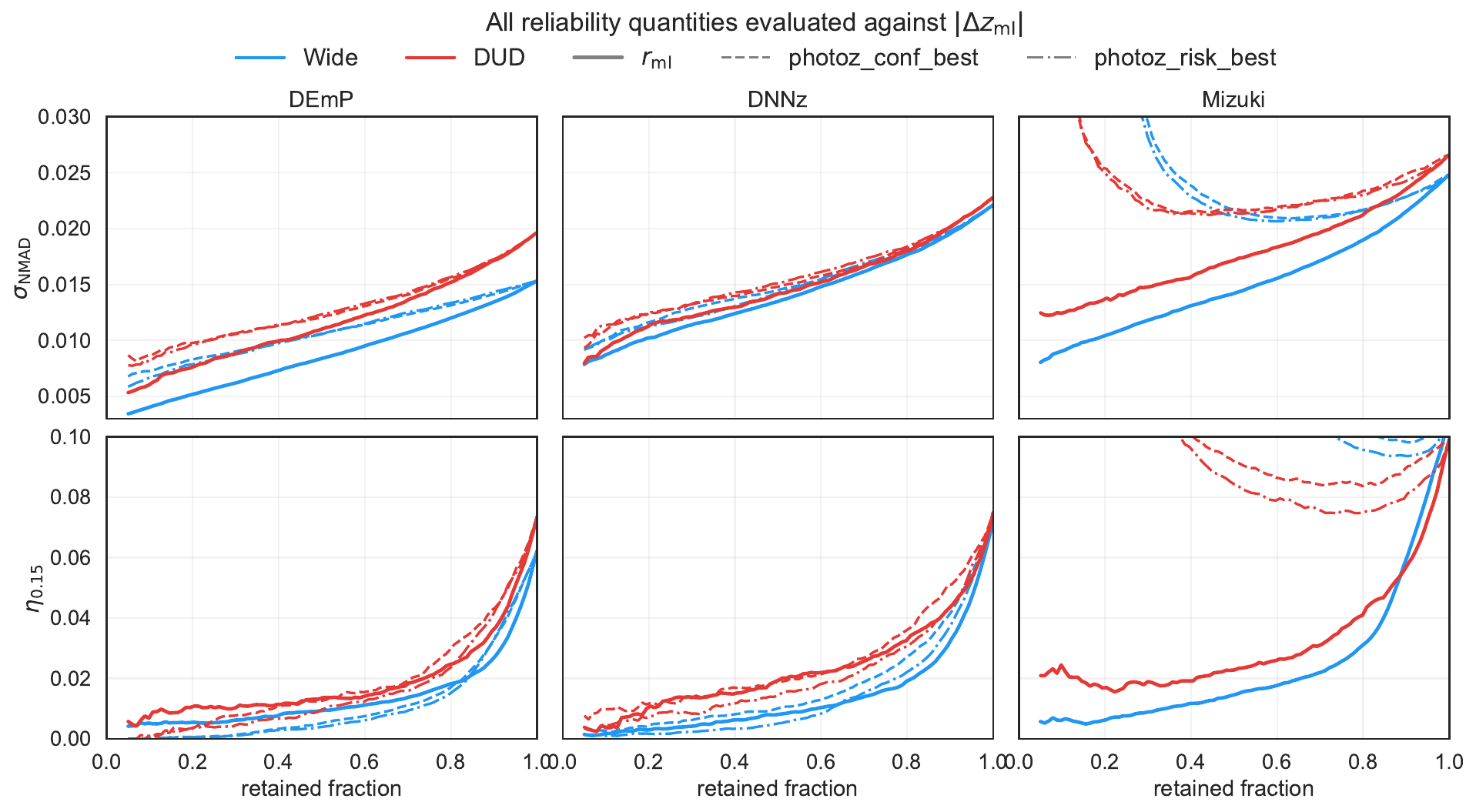}
\includegraphics[trim={0 0 0 2.4cm}, clip, width=.9 \linewidth]{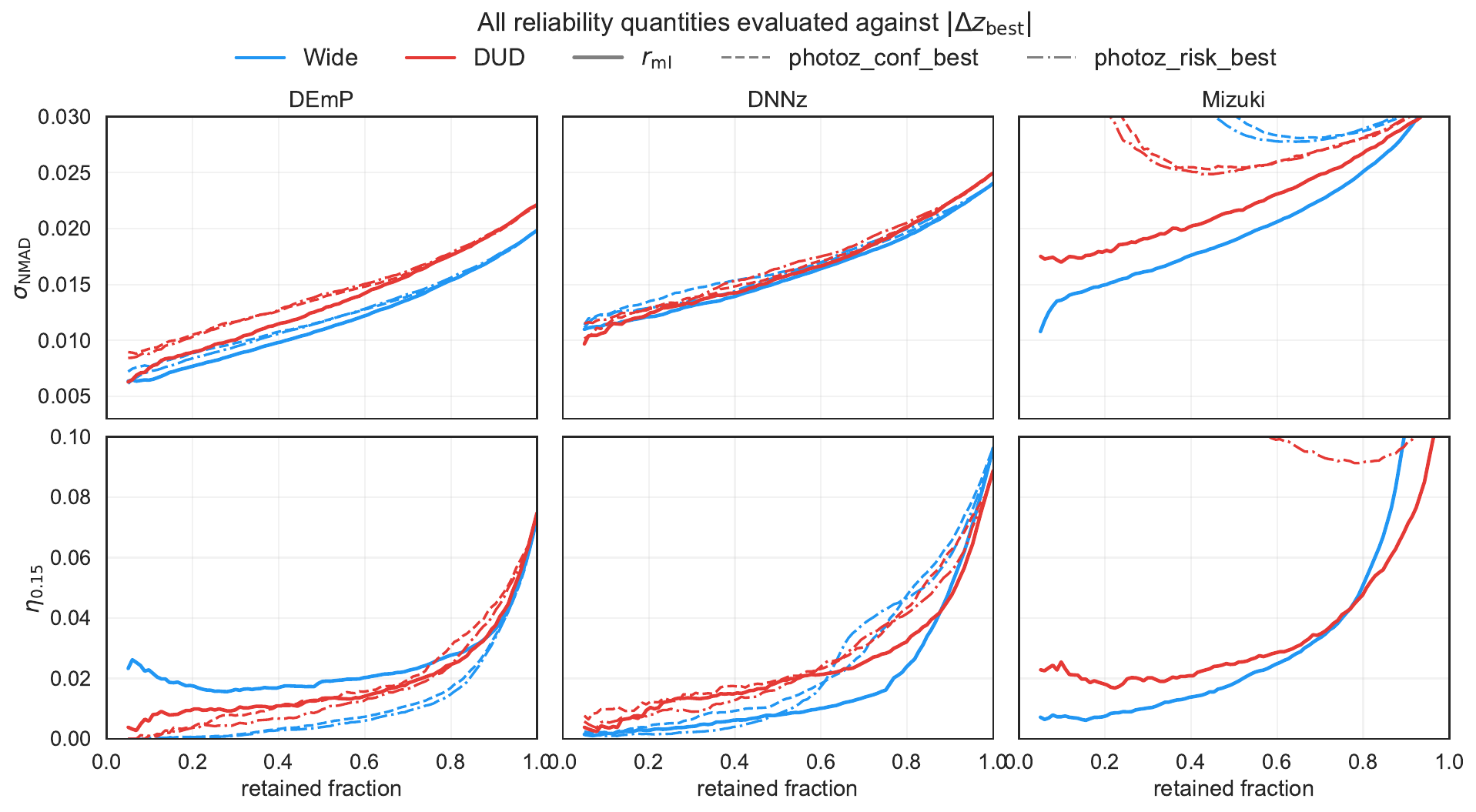}
\caption{Controlled retained-fraction comparison in which all selectors are evaluated using $|\Delta z_{\rm ml}|$ (top) and $|\Delta z_{\rm best}|$ (bottom).\\
Alt text: Two stacked grids of six line plots, with three columns and two rows each. Colors distinguish survey layers; line styles distinguish selectors.}
\label{fig:common-residual}
\end{figure*}

Figure~\ref{fig:common-residual} isolates the effect of the reliability selector by holding the test objects and residual definition fixed within each pipeline-layer comparison. The upper half uses $|\Delta z_{\rm ml}|$ and the lower half uses $|\Delta z_{\rm best}|$. For a fixed configuration and residual, the three selector curves meet at retained fraction $f=1$, when no galaxies have been removed. As $f$ decreases, lower scatter or outlier-fraction curves indicate more effective filtering.

The areas under these curves summarize the comparison. Across the six configurations and two residual targets, $\rml$ has the lowest ${\rm AUC}_\sigma$ in all 12 cases, but the lowest ${\rm AUC}_\eta$ in only six of 12. Paired-bootstrap intervals favor $\rml$ over both catalog selectors in all 24 scatter-AUC comparisons. For outlier AUC, 14 of 24 pairwise differences favor $\rml$, seven favor a catalog selector, and three are unresolved. A separate Spearman-ranking check favors $\rml$ in 11 of 12 cases; the small exception is DEmP-Wide against $|\Delta z_{\rm best}|$, where \mytt{photoz_conf_best} scores 0.568 versus 0.562 for $\rml$. Thus the scatter-filtering advantage persists with either residual definition, whereas the outlier-filtering result remains pipeline dependent.

\section{Paired-bootstrap assessment}
\label{ap:bootstrap}

We quantify sampling uncertainty with 2,000 paired, object-level bootstrap resamples for each pipeline-layer configuration. The point-estimate comparison uses the  {15\% $\zml$ holdout, whereas the error-scale and filtering comparisons use the 7.5\% final-test subset reserved for reliability evaluation}. Within a replicate, the same sampled object indices are applied to both quantities in every pair; all correlations and AUCs for a given reliability replicate share that draw. We report percentile 95\% confidence intervals for a signed improvement $I$, defined so that $I>0$ always favors \mytt{turboPDZ}: catalog minus \mytt{turboPDZ} for lower-is-better point metrics and AUCs, and \mytt{turboPDZ} minus catalog for correlations.
These intervals quantify finite-test-sample variation, but not uncertainty from spectroscopic selection, hyperparameter selection, network refitting, or transfer to the full photometric catalog.

\begin{table}
\renewcommand{\vale}[1]{#1}
\centering
\footnotesize
\setlength{\tabcolsep}{3pt}
\renewcommand{\arraystretch}{1.05}
\caption{Paired-bootstrap differences $I_m=m(\zbest)-m(\zml)$ for $m=\sigma_{\rm NMAD},\eta_{0.15}$ (the latter in percentage points). Entries give the observed difference and its 95\% interval. Bold red favors $\zml$; black includes zero.}
\label{tab:bootstrap-point}
\begin{tabular}{llrr}
\toprule
Pipeline & Layer & $I_{\sigma}$ & $I_{\eta}$ [pp] \\
\midrule
DEmP & Wide & \textcolor{red}{\textbf{+0.0046 [+0.0045, +0.0048]}} & \textcolor{red}{\textbf{+1.08 [+0.96, +1.19]}} \\
DEmP & DUD & \textcolor{red}{\textbf{+0.0024 [+0.0021, +0.0028]}} & +0.16 [-0.02, +0.33] \\
DNNz & Wide & \textcolor{red}{\textbf{+0.0020 [+0.0018, +0.0021]}} & \textcolor{red}{\textbf{+2.43 [+2.27, +2.58]}} \\
DNNz & DUD & \textcolor{red}{\textbf{+0.0023 [+0.0019, +0.0027]}} & \textcolor{red}{\textbf{+1.34 [+1.09, +1.59]}} \\
Mizuki & Wide & \textcolor{red}{\textbf{+0.0090 [+0.0087, +0.0094]}} & \textcolor{red}{\textbf{+7.37 [+7.16, +7.58]}} \\
Mizuki & DUD & \textcolor{red}{\textbf{+0.0058 [+0.0052, +0.0064]}} & \textcolor{red}{\textbf{+2.12 [+1.80, +2.45]}} \\
\bottomrule
\end{tabular}

\end{table}

\begin{table}
\renewcommand{\vale}[1]{#1}
\centering
\footnotesize
\setlength{\tabcolsep}{1pt}
\renewcommand{\arraystretch}{1.0}
\caption{Paired-bootstrap correlation improvements, $I_\rho=\rho(\sml)-\rho(s_{\rm best})$ and $I_r=r(\sml)-r(s_{\rm best})$. ``Native'' pairs each error-scale estimator with its own residual; the controlled rows use the stated common residual. Bold red favors $\sml$ and bold blue favors $s_{\rm best}$ when the interval excludes zero; black is unresolved.}
\label{tab:bootstrap-correlations}
\begin{tabular}{lllrr}
\toprule
Pipel. & Layer & Residual & $I_{\rho}$ & $I_r$ \\
\midrule
DEmP & Wide & native & +0.005 [-0.003, +0.012] & +0.003 [-0.030, +0.039] \\
DEmP & Wide & $|\Delta z_{\rm ml}|$ & \textcolor{red}{\textbf{+0.165 [+0.158, +0.172]}} & \textcolor{red}{\textbf{+0.087 [+0.054, +0.122]}} \\
DEmP & Wide & $|\Delta z_{\rm best}|$ & \textcolor{blue}{\textbf{-0.022 [-0.029, -0.016]}} & +0.015 [-0.019, +0.050] \\
\addlinespace
DEmP & DUD & native & \textcolor{red}{\textbf{+0.023 [+0.008, +0.037]}} & -0.007 [-0.067, +0.073] \\
DEmP & DUD & $|\Delta z_{\rm ml}|$ & \textcolor{red}{\textbf{+0.059 [+0.050, +0.069]}} & +0.030 [-0.027, +0.105] \\
DEmP & DUD & $|\Delta z_{\rm best}|$ & \textcolor{red}{\textbf{+0.029 [+0.020, +0.038]}} & +0.047 [-0.024, +0.138] \\
\addlinespace
DNNz & Wide & native & \textcolor{red}{\textbf{+0.072 [+0.064, +0.079]}} & \textcolor{red}{\textbf{+0.091 [+0.050, +0.130]}} \\
DNNz & Wide & $|\Delta z_{\rm ml}|$ & \textcolor{red}{\textbf{+0.079 [+0.074, +0.084]}} & \textcolor{red}{\textbf{+0.074 [+0.031, +0.111]}} \\
DNNz & Wide & $|\Delta z_{\rm best}|$ & \textcolor{red}{\textbf{+0.075 [+0.070, +0.080]}} & \textcolor{red}{\textbf{+0.150 [+0.129, +0.175]}} \\
\addlinespace
DNNz & DUD & native & \textcolor{red}{\textbf{+0.037 [+0.022, +0.052]}} & +0.046 [-0.002, +0.102] \\
DNNz & DUD & $|\Delta z_{\rm ml}|$ & \textcolor{red}{\textbf{+0.052 [+0.042, +0.061]}} & \textcolor{red}{\textbf{+0.073 [+0.030, +0.121]}} \\
DNNz & DUD & $|\Delta z_{\rm best}|$ & \textcolor{red}{\textbf{+0.039 [+0.028, +0.048]}} & \textcolor{red}{\textbf{+0.092 [+0.026, +0.159]}} \\
\addlinespace
Mizuki & Wide & native & \textcolor{red}{\textbf{+0.628 [+0.614, +0.644]}} & \textcolor{red}{\textbf{+0.458 [+0.438, +0.479]}} \\
Mizuki & Wide & $|\Delta z_{\rm ml}|$ & \textcolor{red}{\textbf{+0.590 [+0.576, +0.604]}} & \textcolor{red}{\textbf{+0.343 [+0.319, +0.366]}} \\
Mizuki & Wide & $|\Delta z_{\rm best}|$ & \textcolor{red}{\textbf{+0.648 [+0.634, +0.663]}} & \textcolor{red}{\textbf{+0.688 [+0.669, +0.709]}} \\
\addlinespace
Mizuki & DUD & native & \textcolor{red}{\textbf{+0.334 [+0.306, +0.359]}} & \textcolor{red}{\textbf{+0.140 [+0.092, +0.189]}} \\
Mizuki & DUD & $|\Delta z_{\rm ml}|$ & \textcolor{red}{\textbf{+0.340 [+0.315, +0.363]}} & \textcolor{red}{\textbf{+0.123 [+0.074, +0.176]}} \\
Mizuki & DUD & $|\Delta z_{\rm best}|$ & \textcolor{red}{\textbf{+0.307 [+0.282, +0.332]}} & \textcolor{red}{\textbf{+0.158 [+0.110, +0.205]}} \\
\addlinespace
\bottomrule
\end{tabular}

\end{table}

\begin{table}
\renewcommand{\vale}[1]{#1}
\centering
\scriptsize
\setlength{\tabcolsep}{1pt}
\renewcommand{\arraystretch}{1.0}
\caption{Paired-bootstrap improvements for native and controlled retained-fraction AUC comparisons. For each catalog comparator, $I_{\rm AUC}={\rm AUC}_{\rm catalog}-{\rm AUC}_{\rml}$, so positive values favor $\rml$. In native rows, $\rml$ accompanies $\zml$ and the catalog selector accompanies $\zbest$; in controlled rows, all selectors use the listed common residual. Bold red favors $\rml$ and bold blue favors the catalog selector when the interval excludes zero; black is unresolved.}
\label{tab:bootstrap-filter-auc}
\begin{tabular}{llllrr}
\toprule
Pipeline & Layer & Residual & Comparator & $I_{{\rm AUC},\sigma}$ & $I_{{\rm AUC},\eta}$ \\
\midrule
DEmP & Wide & native & \texttt{conf\_best} & \textcolor{red}{\textbf{+0.34 [+0.33, +0.35]}} & \textcolor{blue}{\textbf{-0.20 [-0.28, -0.12]}} \\
DEmP & Wide & native & \texttt{risk\_best} & \textcolor{red}{\textbf{+0.33 [+0.32, +0.34]}} & \textcolor{blue}{\textbf{-0.28 [-0.36, -0.20]}} \\
DEmP & Wide & $|\Delta z_{\rm ml}|$ & \texttt{conf\_best} & \textcolor{red}{\textbf{+0.19 [+0.19, +0.20]}} & \textcolor{blue}{\textbf{-0.23 [-0.31, -0.15]}} \\
DEmP & Wide & $|\Delta z_{\rm ml}|$ & \texttt{risk\_best} & \textcolor{red}{\textbf{+0.19 [+0.18, +0.19]}} & \textcolor{blue}{\textbf{-0.29 [-0.37, -0.21]}} \\
DEmP & Wide & $|\Delta z_{\rm best}|$ & \texttt{conf\_best} & \textcolor{red}{\textbf{+0.06 [+0.05, +0.07]}} & \textcolor{blue}{\textbf{-1.19 [-1.35, -1.04]}} \\
DEmP & Wide & $|\Delta z_{\rm best}|$ & \texttt{risk\_best} & \textcolor{red}{\textbf{+0.05 [+0.04, +0.06]}} & \textcolor{blue}{\textbf{-1.27 [-1.43, -1.12]}} \\
\addlinespace
DEmP & DUD & native & \texttt{conf\_best} & \textcolor{red}{\textbf{+0.25 [+0.23, +0.28]}} & +0.02 [-0.15, +0.17] \\
DEmP & DUD & native & \texttt{risk\_best} & \textcolor{red}{\textbf{+0.26 [+0.23, +0.29]}} & \textcolor{blue}{\textbf{-0.23 [-0.40, -0.07]}} \\
DEmP & DUD & $|\Delta z_{\rm ml}|$ & \texttt{conf\_best} & \textcolor{red}{\textbf{+0.11 [+0.09, +0.12]}} & -0.01 [-0.16, +0.14] \\
DEmP & DUD & $|\Delta z_{\rm ml}|$ & \texttt{risk\_best} & \textcolor{red}{\textbf{+0.11 [+0.09, +0.13]}} & \textcolor{blue}{\textbf{-0.24 [-0.41, -0.10]}} \\
DEmP & DUD & $|\Delta z_{\rm best}|$ & \texttt{conf\_best} & \textcolor{red}{\textbf{+0.08 [+0.07, +0.09]}} & +0.03 [-0.12, +0.18] \\
DEmP & DUD & $|\Delta z_{\rm best}|$ & \texttt{risk\_best} & \textcolor{red}{\textbf{+0.08 [+0.07, +0.10]}} & \textcolor{blue}{\textbf{-0.21 [-0.37, -0.07]}} \\
\addlinespace
DNNz & Wide & native & \texttt{conf\_best} & \textcolor{red}{\textbf{+0.25 [+0.23, +0.27]}} & \textcolor{red}{\textbf{+1.10 [+1.01, +1.19]}} \\
DNNz & Wide & native & \texttt{risk\_best} & \textcolor{red}{\textbf{+0.23 [+0.21, +0.24]}} & \textcolor{red}{\textbf{+0.95 [+0.86, +1.04]}} \\
DNNz & Wide & $|\Delta z_{\rm ml}|$ & \texttt{conf\_best} & \textcolor{red}{\textbf{+0.08 [+0.08, +0.09]}} & \textcolor{red}{\textbf{+0.36 [+0.29, +0.42]}} \\
DNNz & Wide & $|\Delta z_{\rm ml}|$ & \texttt{risk\_best} & \textcolor{red}{\textbf{+0.07 [+0.06, +0.07]}} & -0.00 [-0.07, +0.06] \\
DNNz & Wide & $|\Delta z_{\rm best}|$ & \texttt{conf\_best} & \textcolor{red}{\textbf{+0.07 [+0.07, +0.08]}} & \textcolor{red}{\textbf{+0.76 [+0.68, +0.83]}} \\
DNNz & Wide & $|\Delta z_{\rm best}|$ & \texttt{risk\_best} & \textcolor{red}{\textbf{+0.05 [+0.05, +0.06]}} & \textcolor{red}{\textbf{+0.61 [+0.54, +0.69]}} \\
\addlinespace
DNNz & DUD & native & \texttt{conf\_best} & \textcolor{red}{\textbf{+0.18 [+0.15, +0.21]}} & \textcolor{red}{\textbf{+0.53 [+0.33, +0.73]}} \\
DNNz & DUD & native & \texttt{risk\_best} & \textcolor{red}{\textbf{+0.21 [+0.18, +0.24]}} & \textcolor{red}{\textbf{+0.28 [+0.08, +0.48]}} \\
DNNz & DUD & $|\Delta z_{\rm ml}|$ & \texttt{conf\_best} & \textcolor{red}{\textbf{+0.07 [+0.05, +0.08]}} & \textcolor{red}{\textbf{+0.20 [+0.02, +0.37]}} \\
DNNz & DUD & $|\Delta z_{\rm ml}|$ & \texttt{risk\_best} & \textcolor{red}{\textbf{+0.09 [+0.07, +0.10]}} & \textcolor{blue}{\textbf{-0.22 [-0.39, -0.06]}} \\
DNNz & DUD & $|\Delta z_{\rm best}|$ & \texttt{conf\_best} & \textcolor{red}{\textbf{+0.03 [+0.01, +0.05]}} & \textcolor{red}{\textbf{+0.45 [+0.27, +0.64]}} \\
DNNz & DUD & $|\Delta z_{\rm best}|$ & \texttt{risk\_best} & \textcolor{red}{\textbf{+0.06 [+0.04, +0.07]}} & \textcolor{red}{\textbf{+0.20 [+0.01, +0.38]}} \\
\addlinespace
Mizuki & Wide & native & \texttt{conf\_best} & \textcolor{red}{\textbf{+5.17 [+5.01, +5.35]}} & \textcolor{red}{\textbf{+27.07 [+26.68, +27.49]}} \\
Mizuki & Wide & native & \texttt{risk\_best} & \textcolor{red}{\textbf{+4.97 [+4.81, +5.13]}} & \textcolor{red}{\textbf{+26.69 [+26.30, +27.14]}} \\
Mizuki & Wide & $|\Delta z_{\rm ml}|$ & \texttt{conf\_best} & \textcolor{red}{\textbf{+1.95 [+1.87, +2.04]}} & \textcolor{red}{\textbf{+12.84 [+12.43, +13.27]}} \\
Mizuki & Wide & $|\Delta z_{\rm ml}|$ & \texttt{risk\_best} & \textcolor{red}{\textbf{+2.03 [+1.94, +2.12]}} & \textcolor{red}{\textbf{+12.63 [+12.20, +13.06]}} \\
Mizuki & Wide & $|\Delta z_{\rm best}|$ & \texttt{conf\_best} & \textcolor{red}{\textbf{+4.67 [+4.51, +4.84]}} & \textcolor{red}{\textbf{+25.77 [+25.41, +26.16]}} \\
Mizuki & Wide & $|\Delta z_{\rm best}|$ & \texttt{risk\_best} & \textcolor{red}{\textbf{+4.46 [+4.31, +4.62]}} & \textcolor{red}{\textbf{+25.40 [+25.03, +25.80]}} \\
\addlinespace
Mizuki & DUD & native & \texttt{conf\_best} & \textcolor{red}{\textbf{+1.95 [+1.65, +2.31]}} & \textcolor{red}{\textbf{+12.50 [+11.39, +13.59]}} \\
Mizuki & DUD & native & \texttt{risk\_best} & \textcolor{red}{\textbf{+1.81 [+1.51, +2.17]}} & \textcolor{red}{\textbf{+11.43 [+10.38, +12.50]}} \\
Mizuki & DUD & $|\Delta z_{\rm ml}|$ & \texttt{conf\_best} & \textcolor{red}{\textbf{+0.90 [+0.74, +1.09]}} & \textcolor{red}{\textbf{+8.66 [+7.76, +9.56]}} \\
Mizuki & DUD & $|\Delta z_{\rm ml}|$ & \texttt{risk\_best} & \textcolor{red}{\textbf{+0.85 [+0.70, +1.02]}} & \textcolor{red}{\textbf{+7.97 [+7.11, +8.88]}} \\
Mizuki & DUD & $|\Delta z_{\rm best}|$ & \texttt{conf\_best} & \textcolor{red}{\textbf{+1.49 [+1.20, +1.84]}} & \textcolor{red}{\textbf{+12.06 [+11.01, +13.14]}} \\
Mizuki & DUD & $|\Delta z_{\rm best}|$ & \texttt{risk\_best} & \textcolor{red}{\textbf{+1.35 [+1.06, +1.70]}} & \textcolor{red}{\textbf{+10.99 [+9.97, +12.05]}} \\
\addlinespace
\bottomrule
\end{tabular}

\end{table}

Table~\ref{tab:bootstrap-point} supports the $\sigma_{\rm NMAD}$ improvement in all six configurations and the $\eta_{0.15}$ improvement in five; DEmP-DUD is unresolved. The paired MAE intervals also favor $\zml$ in all six configurations but are omitted from the table for compactness.
Tables~\ref{tab:bootstrap-correlations} and \ref{tab:bootstrap-filter-auc} show where the numerical rankings are robust. In the native error-scale comparison, eight of 12 correlation differences support $\sml$ and four remain unresolved; none supports $s_{\rm best}$. In the controlled comparison, 20 of 24 support $\sml$, three Pearson differences remain unresolved, and the DEmP-Wide Spearman comparison against $|\Delta z_{\rm best}|$ supports $s_{\rm best}$. For filtering, every native and controlled ${\rm AUC}_\sigma$ pair supports $\rml$. The ${\rm AUC}_\eta$ results are more configuration dependent: eight of 12 native pairs and 14 of 24 controlled pairs support $\rml$, while three native and seven controlled pairs support a catalog selector; the remaining one native and three controlled comparisons are unresolved.

\section{Hyperparameter optimization}
\label{ap:hp}

 {The \texttt{Optuna} searches were performed independently for each catalog/layer pair on the magnitude- and redshift-limited sample, before imposing \texttt{specz\_flag\_homogeneous=True}. Each trial used five-fold stratified cross-validation within its search training pool, with a TPE univariate sampler and percentile pruning. For $\zml$, that pool contained 85\% of the original sample, with a 15\% holdout. For $\sml$, the original training/calibration/test fractions were 80/10/10.}

 {After the secure-redshift selection, the selected network hyperparameters and epoch budgets are retained without a new search. The 85/15 partition is constructed from the secure sample, separately from the search partitions, and shared by the point-estimate fits and the reliability analysis. The evaluation objects are excluded from fitting and calibration, but some contributed to the earlier hyperparameter searches. Restricting the final-test sample to objects absent from both search pools preserves the native scatter-filtering advantage of $\rml$ in all six configurations. Paired-bootstrap 95\% intervals support this advantage over both catalog selectors in five configurations; DNNz-DUD remains unresolved on its smaller independent subset.}

Within each fold, training uses a maximum of 150 epochs, early stopping on validation loss (patience $=18$, restoring the best weights), and \texttt{ReduceLROnPlateau} (factor $=0.5$, patience $=6$, minimum learning rate $=10^{-5}$). The \texttt{Optuna} setup for $\zml$ uses 600 total complete trials, 25 TPE startup trials, and a percentile pruner at the 85th percentile (25 startup trials, 2 warmup steps, pruning interval $=1$).
The \texttt{Optuna} setup for $\sml$ uses similar settings but a smaller budget of 200 complete trials; the selected architectures alone do not establish optimizer convergence.
 {Tables~\ref{tab:best_hp} and \ref{tab:best_hp_sigmaml} list the selected hyperparameters reused for each pipeline and survey layer.}
The explored hyperparameter space is summarized in Table~\ref{tab:hp}. Diagnostic plots for the optimization process, such as trial history and hyperparameter importance, will accompany the public-library release at \href{https://github.com/valerio-marra/turboPDZ}{github.com/valerio-marra/turboPDZ}.

The importance analysis for the $\zml$ optimization reveals a consistent pattern across all six pipeline-layer combinations. The learning rate is by far the most influential hyperparameter in every case, accounting for roughly half or more of the total importance budget. This dominance is expected: the learning rate directly controls the convergence dynamics of the AdamW optimizer and strongly interacts with the early-stopping and learning-rate reduction schedule.
 The number of hidden layers is consistently the second most important parameter, followed by the explained variance fraction for PCA compression.
 The activation function, batch size, dropout rate, and weight decay each contribute modestly. Individual per-layer unit counts carry very little importance on their own, indicating that the optimizer is relatively insensitive to the precise width of each layer as long as the total capacity (set by the number of layers) is adequate.
 Overall, the importance rankings are remarkably stable across the three photo-$z$ pipelines and the two survey layers, suggesting that the $\zml$ HPO search space is well behaved and that the dominant optimization levers are universal rather than pipeline-specific.

 The $\sml$ optimization shows a less uniform picture. For DEmP and DNNz, the learning rate remains the dominant parameter (importance 0.73--0.82), broadly matching the $\zml$ pattern. For Mizuki, however, the landscape shifts: in the DUD layer, weight decay dominates (0.65) with the learning rate contributing negligibly (0.01), while in the Wide layer, the dropout rate is the leading parameter (0.29) and batch size is second (0.18).
 This suggests that, for template-fitting PDZs, regularization plays a larger role in the $\sml$ optimization, possibly reflecting the higher noise level of the log-error target for this pipeline.

We additionally tested alternative model families (XGBoost, LightGBM, CatBoost), alternative losses and activations, and optional redshift weighting; none produced a robust improvement over the adopted composite-objective neural pipeline.

\begin{table}
\centering
\setlength{\tabcolsep}{10pt}
\renewcommand{\arraystretch}{1.4}
\caption{Search space used in the \texttt{Optuna} optimization.}
\label{tab:hp}
\begin{tabular}{ll}
\toprule
Hyperparameter & Range \\
\midrule
Explained variance (PCA) & $[0.50,\,0.90]$ \\
Batch size & $\{64,\,128,\,256,\,512\}$ \\
Number of hidden layers & $[3,\,7]$ \\
Units per hidden layer & $[192,\,1024]$ (step 64) \\
Dropout rate & $[0.10,\,0.35]$ \\
Learning rate & $[10^{-4},\, 10^{-2}]$ (log-uniform) \\
Weight decay & $[10^{-6},\,5 \times 10^{-3}]$ (log-uniform) \\
Activation & \{ReLU, ELU, Swish\} \\
\bottomrule
\end{tabular}
\end{table}


\bibliographystyle{aaArxivDoi}
\bibliography{biblio}

\begin{thebibliography}{24}
\expandafter\ifx\csname natexlab\endcsname\relax\def\natexlab#1{#1}\fi

\bibitem[{Aihara {et~al.}(2018)}]{Aihara:2017paw}
Aihara, H. {et~al.} 2018, \href{https://doi.org/10.1093/pasj/psx066}{Publ. Astron. Soc. Jap.}, 70, S4, [\href{https://arxiv.org/abs/1704.05858}{1704.05858}].

\bibitem[{Aihara {et~al.}(2022)}]{Aihara:2021jwb}
Aihara, H. {et~al.} 2022, \href{https://doi.org/10.1093/pasj/psab122}{Publ. Astron. Soc. Jap.}, 74, 247, [\href{https://arxiv.org/abs/2108.13045}{2108.13045}].

\bibitem[{{Akiba} {et~al.}(2019){Akiba}, {Sano}, {Yanase}, {Ohta}, \& {Koyama}}]{2019arXiv190710902A}
{Akiba}, T., {Sano}, S., {Yanase}, T., {Ohta}, T., \& {Koyama}, M. 2019, \href{https://doi.org/10.48550/arXiv.1907.10902}{arXiv e-prints}, arXiv:1907.10902, [\href{https://arxiv.org/abs/1907.10902}{1907.10902}].

\bibitem[{Bechtol {et~al.}(2026)}]{DES:2025key}
Bechtol, K. {et~al.} 2026, \href{https://doi.org/10.3847/1538-4365/ae18d3}{Astrophys. J. Suppl.}, 282, 62, [\href{https://arxiv.org/abs/2501.05739}{2501.05739}].

\bibitem[{{Ben{\'\i}tez}(2000)}]{2000ApJ...536..571B}
{Ben{\'\i}tez}, N. 2000, \href{https://doi.org/10.1086/308947}{\apj}, 536, 571, [\href{https://arxiv.org/abs/astro-ph/9811189}{astro-ph/9811189}].

\bibitem[{Bonoli {et~al.}(2021)}]{Bonoli:2020ciz}
Bonoli, S. {et~al.} 2021, \href{https://doi.org/10.1051/0004-6361/202038841}{Astron. Astrophys.}, 653, A31, [\href{https://arxiv.org/abs/2007.01910}{2007.01910}].

\bibitem[{Dalal {et~al.}(2023)}]{Dalal:2023olq}
Dalal, R. {et~al.} 2023, \href{https://doi.org/10.1103/PhysRevD.108.123519}{Phys. Rev. D}, 108, 123519, [\href{https://arxiv.org/abs/2304.00701}{2304.00701}].

\bibitem[{Desprez {et~al.}(2020)}]{Euclid:2020gbk}
Desprez, G. {et~al.} 2020, \href{https://doi.org/10.1051/0004-6361/202039403}{Astron. Astrophys.}, 644, A31, [\href{https://arxiv.org/abs/2009.12112}{2009.12112}].

\bibitem[{Ferreira \& Reis(2025)}]{Ferreira:2025mnb}
Ferreira, P.~S. \& Reis, R. R.~R. 2025, \href{https://doi.org/10.1088/1475-7516/2025/07/008}{JCAP}, 07, 008, [\href{https://arxiv.org/abs/2501.04118}{2501.04118}].

\bibitem[{Hern{\'a}n-Caballero {et~al.}(2021)}]{Hernan-Caballero:2021aig}
Hern{\'a}n-Caballero, A. {et~al.} 2021, \href{https://doi.org/10.1051/0004-6361/202141236}{Astron. Astrophys.}, 654, A101, [\href{https://arxiv.org/abs/2108.03271}{2108.03271}].

\bibitem[{{Hsieh} \& {Yee}(2014)}]{2014ApJ...792..102H}
{Hsieh}, B.~C. \& {Yee}, H.~K.~C. 2014, \href{https://doi.org/10.1088/0004-637X/792/2/102}{\apj}, 792, 102, [\href{https://arxiv.org/abs/1407.5151}{1407.5151}].

\bibitem[{{Jouvel} {et~al.}(2014){Jouvel}, {Host}, {Lahav}, {Seitz}, {Molino}, {Coe}, {Postman}, {Moustakas}, {Ben{\`\i}tez}, {Rosati}, {Balestra}, {Grillo}, {Bradley}, {Fritz}, {Kelson}, {Koekemoer}, {Lemze}, {Medezinski}, {Mercurio}, {Moustakas}, {Nonino}, {Scodeggio}, {Zheng}, {Zitrin}, {Bartelmann}, {Bouwens}, {Broadhurst}, {Donahue}, {Ford}, {Graves}, {Infante}, {Jimenez-Teja}, {Lazkoz}, {Melchior}, {Meneghetti}, {Merten}, {Ogaz}, \& {Umetsu}}]{2014A&A...562A..86J}
{Jouvel}, S., {Host}, O., {Lahav}, O., {et~al.} 2014, \href{https://doi.org/10.1051/0004-6361/201322419}{\aap}, 562, A86, [\href{https://arxiv.org/abs/1308.0063}{1308.0063}].

\bibitem[{Mellier {et~al.}(2025)}]{Euclid:2024yrr}
Mellier, Y. {et~al.} 2025, \href{https://doi.org/10.1051/0004-6361/202450810}{Astron. Astrophys.}, 697, A1, [\href{https://arxiv.org/abs/2405.13491}{2405.13491}].

\bibitem[{{Newman} \& {Gruen}(2022)}]{2022ARA&A..60..363N}
{Newman}, J.~A. \& {Gruen}, D. 2022, \href{https://doi.org/10.1146/annurev-astro-032122-014611}{\araa}, 60, 363, [\href{https://arxiv.org/abs/2206.13633}{2206.13633}].

\bibitem[{Nishizawa {et~al.}(2021)Nishizawa, Hsieh, \& Tanaka}]{Nishizawa:pdr3}
Nishizawa, A.~J., Hsieh, B.-C., \& Tanaka, M. 2021, Photometric Redshifts for the Hyper Suprime-Cam Subaru Strategic Program Data Release 3, HSC-SSP website, PDR3, available at \url{https://hsc-release.mtk.nao.ac.jp/doc/wp-content/uploads/2022/08/pdr3_photoz.pdf}

\bibitem[{{Nishizawa} {et~al.}(2020){Nishizawa}, {Hsieh}, {Tanaka}, \& {Takata}}]{2020arXiv200301511N}
{Nishizawa}, A.~J., {Hsieh}, B.-C., {Tanaka}, M., \& {Takata}, T. 2020, \href{https://doi.org/10.48550/arXiv.2003.01511}{arXiv e-prints}, arXiv:2003.01511, [\href{https://arxiv.org/abs/2003.01511}{2003.01511}].

\bibitem[{{Rau} {et~al.}(2015){Rau}, {Seitz}, {Brimioulle}, {Frank}, {Friedrich}, {Gruen}, \& {Hoyle}}]{2015MNRAS.452.3710R}
{Rau}, M.~M., {Seitz}, S., {Brimioulle}, F., {et~al.} 2015, \href{https://doi.org/10.1093/mnras/stv1567}{\mnras}, 452, 3710, [\href{https://arxiv.org/abs/1503.08215}{1503.08215}].

\bibitem[{{Salvato} {et~al.}(2019){Salvato}, {Ilbert}, \& {Hoyle}}]{2019NatAs...3..212S}
{Salvato}, M., {Ilbert}, O., \& {Hoyle}, B. 2019, \href{https://doi.org/10.1038/s41550-018-0478-0}{Nature Astronomy}, 3, 212, [\href{https://arxiv.org/abs/1805.12574}{1805.12574}].

\bibitem[{Schmidt {et~al.}(2020)}]{LSSTDarkEnergyScience:2020nwm}
Schmidt, S.~J. {et~al.} 2020, \href{https://doi.org/10.1093/mnras/staa2799}{Mon. Not. Roy. Astron. Soc.}, 499, 1587, [\href{https://arxiv.org/abs/2001.03621}{2001.03621}].

\bibitem[{Sevilla-Noarbe {et~al.}(2021)}]{DES:2020aks}
Sevilla-Noarbe, I. {et~al.} 2021, \href{https://doi.org/10.3847/1538-4365/abeb66}{Astrophys. J. Suppl.}, 254, 24, [\href{https://arxiv.org/abs/2011.03407}{2011.03407}].

\bibitem[{{Tanaka}(2015)}]{2015ApJ...801...20T}
{Tanaka}, M. 2015, \href{https://doi.org/10.1088/0004-637X/801/1/20}{\apj}, 801, 20, [\href{https://arxiv.org/abs/1501.02047}{1501.02047}].

\bibitem[{Tanaka {et~al.}(2018)Tanaka, Coupon, Hsieh, Mineo, Nishizawa, Speagle, Furusawa, Miyazaki, \& Murayama}]{Tanaka:2017lit}
Tanaka, M., Coupon, J., Hsieh, B.-C., {et~al.} 2018, \href{https://doi.org/10.1093/pasj/psx077}{Publ. Astron. Soc. Jap.}, 70, S9, [\href{https://arxiv.org/abs/1704.05988}{1704.05988}].

\bibitem[{{Teixeira} {et~al.}(2024){Teixeira}, {Bom}, {Santana-Silva}, {Fraga}, {Darc}, {Teixeira}, {Wu}, {Ferguson}, {Mart{\'\i}nez-V{\'a}zquez}, {Riley}, {Drlica-Wagner}, {Choi}, {Mutlu-Pakdil}, {Pace}, {Sakowska}, \& {Stringfellow}}]{2024A&C....4900886T}
{Teixeira}, G., {Bom}, C.~R., {Santana-Silva}, L., {et~al.} 2024, \href{https://doi.org/10.1016/j.ascom.2024.100886}{Astronomy and Computing}, 49, 100886, [\href{https://arxiv.org/abs/2408.15243}{2408.15243}].

\bibitem[{{The LSST Dark Energy Science Collaboration} {et~al.}(2018){The LSST Dark Energy Science Collaboration}, {Mandelbaum}, {Eifler}, {Hlo{\v{z}}ek}, {Collett}, {Gawiser}, {Scolnic}, {Alonso}, {Awan}, {Biswas}, {Blazek}, {Burchat}, {Chisari}, {Dell'Antonio}, {Digel}, {Frieman}, {Goldstein}, {Hook}, {Ivezi{\'c}}, {Kahn}, {Kamath}, {Kirkby}, {Kitching}, {Krause}, {Leget}, {Marshall}, {Meyers}, {Miyatake}, {Newman}, {Nichol}, {Rykoff}, {Sanchez}, {Slosar}, {Sullivan}, \& {Troxel}}]{2018arXiv180901669T}
{The LSST Dark Energy Science Collaboration}, {Mandelbaum}, R., {Eifler}, T., {et~al.} 2018, \href{https://doi.org/10.48550/arXiv.1809.01669}{arXiv e-prints}, arXiv:1809.01669, [\href{https://arxiv.org/abs/1809.01669}{1809.01669}].

\end{thebibliography}

\end{document}